\documentclass[manuscript, screen]{acmart}

\usepackage{graphicx}
\usepackage{textcomp}
\usepackage{multirow}
\usepackage{tikz}
\usepackage{circledsteps}
\usepackage{listings}
\usepackage{xcolor}
\usepackage{colortbl}
\usepackage[export]{adjustbox}
\usepackage[utf8]{inputenc}
\usepackage[english=american]{csquotes}
\usepackage{tabularx}
\usepackage{subcaption}
\usepackage{caption}
\usepackage{tcolorbox}
\definecolor{aliceblue}{rgb}{0.94, 0.97, 1.0}
\usepackage{comment}
\usepackage{url}
\usepackage{balance}
\usepackage[ruled,vlined]{algorithm2e}
\usepackage{graphicx}
\usepackage{footnote}
\usepackage[perpage]{footmisc}
\usepackage{bbm}
\usepackage{rotating}

\usepackage{soul}

\usepackage{tcolorbox}

\usepackage{listings}
\definecolor{codegreen}{rgb}{0,0.6,0}
\definecolor{codegray}{rgb}{0.5,0.5,0.5}
\definecolor{codepurple}{rgb}{0.58,0,0.82}
\definecolor{backcolour}{rgb}{0.9,0.9,0.9}
\definecolor{red}{RGB}{160,0,0}
\lstdefinestyle{mystyle}{
    backgroundcolor=\color{backcolour},   
    commentstyle=\color{codegreen},
    keywordstyle=\color{magenta},
    numberstyle=\tiny\color{codegray},
    stringstyle=\color{codepurple},
    basicstyle=\ttfamily\footnotesize,
    breakatwhitespace=false,         
    breaklines=true,  
    captionpos=b,                    
    keepspaces=true,                 
    numbers=left,                    
    numbersep=5pt,                  
    showspaces=false,                
    showstringspaces=false,
    showtabs=false,                  
    tabsize=2,
}
\AtBeginDocument{%
  }

\setcopyright{acmlicensed}
\copyrightyear{2025}
\acmYear{2025}
\acmDOI{XXXXXXX.XXXXXXX}

\acmJournal{JACM}
\acmVolume{37}
\acmNumber{4}
\acmArticle{111}
\acmMonth{8}

\begin{document}

\title{TaskEval: Assessing Difficulty of Code Generation Tasks for Large Language Models}

\author{Florian Tambon}
\authornote{Work done partly while at Polytechnique Montreal.}
\email{florian.tambon@uni.lu}
\orcid{0000-0001-5593-9400}
\affiliation{%
  \institution{SnT, University of Luxembourg}
  \city{Luxembourg}
  \country{Luxembourg}
}

\author{Amin Nikanjam}
\authornotemark[1]
\email{amin.nikanjam@h-partners.com;amin.nikanjam@polymtl.ca}
\affiliation{%
  \institution{Huawei Distributed Scheduling and Data Engine Lab}
  \city{Toronto}
  \state{Ontario}
  \country{Canada}
}

\author{Cyrine Zid}
\email{cyrine.zid@polymtl.ca}
\affiliation{%
  \institution{Polytechnique Montreal}
  \city{Montreal}
  \state{Quebec}
  \country{Canada}
}

\author{Foutse Khomh}
\email{foutse.khomh@polymtl.ca}
\affiliation{%
  \institution{Polytechnique Montreal}
  \city{Montreal}
  \state{Quebec}
  \country{Canada}
}

\author{Giuliano Antoniol}
\email{giuliano.antoniol@polymtl.ca}
\affiliation{%
  \institution{Polytechnique Montreal}
  \city{Montreal}
  \state{Quebec}
  \country{Canada}
}
\renewcommand{\shortauthors}{Tambon et al.}

\newcommand{\Foutse}[1]{\textcolor{red}{{\it [Foutse: #1]}}}
\newcommand{\Amin}[1]{\textcolor{orange}{{\it [Amin: #1]}}}
\newcommand{\Florian}[1]{\textcolor{magenta}{{\it [Florian: #1]}}}
\newcommand{\Giulio}[1]{\textcolor{codegreen}{{\it [Giulio: #1]}}}
\newcommand{\Cyrine}[1]{\textcolor{blue}{{\it [Cyrine: #1]}}}

\newcommand{\rqone}{How do tasks' characteristics varies across benchmark?}
\newcommand{\rqtwo}{Do tasks' topics affect their characteristics?}
\newcommand{\rqthree}{Do programming constructs affect tasks' characteristics?}
\newcommand{\rqfour}{Do human annotators and CodeLLMs rate tasks' difficulty similarly?}

\begin{abstract}

Large Language Models (LLMs) excel in code-related tasks like code generation, but benchmark evaluations often overlook task characteristics, such as difficulty. Moreover, benchmarks are usually built using tasks described with a single prompt, despite the formulation of prompts having a profound impact on the outcome. This paper introduces a generalist approach, TaskEval, a framework using diverse prompts and Item Response Theory (IRT) to efficiently assess LLMs' capabilities and benchmark task characteristics, improving the understanding of their performance.

Using two code generation benchmarks, \textit{HumanEval}+ and \textit{ClassEval}, as well as 8 code generation LLMs, we show that \textit{TaskEval} is capable of characterising the properties of tasks. Using topic analysis, we identify and analyse the tasks of 17 and 21 topics within the benchmarks. We also cross-analyse tasks' characteristics with programming constructs (e.g., variable assignment, conditions, etc.) used by LLMs, emphasising some patterns with tasks' difficulty. Finally, we conduct a comparison between the difficulty assessment of tasks by human annotators and LLMs. Orthogonal to current benchmarking evaluation efforts, \textit{TaskEval} can assist researchers and practitioners in fostering better assessments of LLMs. The tasks' characteristics can be used to identify shortcomings within existing benchmarks or improve the evaluation of LLMs.

\end{abstract}



\received{20 February 2007}
\received[revised]{12 March 2009}
\received[accepted]{5 June 2009}

\maketitle

\section{Introduction}
Large Language Models (LLMs) have been widely used for various code-related tasks in Software Engineering (SE) \cite{nguyen2022empirical,lemieux2023codamosa,chen2023teaching,MORADIDAKHEL2023111734}. Although the performance of LLMs in generating code for different programming languages, such as Python, Java, and C~\cite{yu2023codereval, Jin23, HumanEvalplus} is promising, the LLM-generated code, like human-written code, is not flawless.  Researchers have already acknowledged that LLM-generated code is prone to bugs~\cite{MORADIDAKHEL2023111734, tambon2024bugs}, and may suffer from vulnerability \cite{toth2024llms,majdinasab2023assessing,klemmer2024using} and quality issues \cite{yeticstiren2023evaluating}. Generally, to assess LLMs on code generation tasks and compare them to each other, well-known benchmarks such as HumanEval \cite{HumanEval,HumanEvalplus}, MBPP \cite{MBPP}, or ClassEval \cite{ClassEval} are used. These benchmarks are obtained by handcrafting programming tasks or mining existing public repositories on GitHub and filtering them. In this way, the code generation capability of LLMs in programming tasks is assessed indirectly at the benchmark level based on general characteristics such as 
granularity level (e.g., function level being less difficult than class level) and through general metrics such as accuracy or Pass@k over the benchmark \cite{HumanEval}. However, this limits the evaluation of the LLMs at the task level as the metric can only report aggregated results over a set of tasks. Moreover, most benchmarks only contain a single prompt for each task within the benchmark and recent studies \cite{MultiPrompts, ReCode} showed that the way prompts are formulated has a high impact on the quality of LLM's output. As such, there is a need for a more fine-grained analysis of the evaluation of the capability of the LLMs at the \textit{task}-level, that goes beyond general aggregation metrics. Better task-level analysis could also pave the way for more advanced and customised benchmarks, as traditional benchmarks have several limitations, such as data contamination, inadequate coverage of LLMs' ability, reliability, and robustness \cite{Livecodebench, mcintosh2024inadequacieslargelanguagemodel, laskar-etal-2024-systematic}.

In the field of test theory, Item Response Theory (IRT) \cite{baker2001basics} was developed precisely to overcome the limitation of considering only an aggregation of the score rather than individual questions. To achieve this, an IRT model estimates a specific ability for each respondent (e.g., a student taking a test or an LLM being evaluated) based on their responses to a set of items (such as questions).
These items have in turn some \textit{characteristics}, for instance, a difficulty level, estimated by the respondents' answers as well. Contrary to using plain metrics such as Accuracy, which do not account for the difference between items (i.e. all LLMs replying correctly on an item vs one LLM replying correctly to it is counted the same for the LLM being correct), IRT will consider each item independently and in conjunction, allowing for a more fine-grained analysis. Moreover, by using IRT and the LLMs themselves, we can go beyond using human assessments as a proxy for concepts such as difficulty. Indeed, given that LLMs' perception of tasks might differ from humans \cite{GPTComp, HumanAttention}, LLMs' grasping of concepts is likely not to be aligned with humans. Recent studies \cite{vania2021comparing, byrd-srivastava-2022-predicting, zhuang2023efficiently} used IRT to assess the characteristics of the questions in a benchmark for an LLM. However, these studies present several limitations. First, they did not focus on code-based benchmarks, which have some differences compared to more classical Natural Language Processing (NLP) tasks. In addition, the assessment is generally conducted via a \textit{single} prompt, which could affect the output of the LLM. Finally, the studies focused mainly on questions for which a single binary (true/false) answer is given. This limits the application of the methods when multiple code generations should be considered, as is the case when using non-deterministic temperature-based sampling for LLMs. Thus, current benchmarks' usage and assessment paint a general, yet incomplete picture, as they fail to provide a fine-grained analysis of individual tasks within a benchmark, accounting for variability in the prompt. To address this issue, we propose \textit{TaskEval}, a framework for the evaluation, identification, and analysis of the difficulty of programming tasks for LLMs.

\textit{TaskEval} is orthogonal to the usage of existing benchmarks and works on top of available benchmarks. \textit{TaskEval} starts by generating a variety of prompts for each task on two well-known code generation benchmarks, HumanEval+ \cite{HumanEval} and ClassEval \cite{ClassEval}, representing multiple possible formulations per task. For each programming task, we craft prompts with different levels of context information and distinctive phrasing to represent diverse possible wording of the task using GPT-4, resulting in 18 different prompts per task. The framework then queries an array of 8 LLMs of varying size, architecture and training process, named CodeLLMs (i.e., YiCoder-1.5B \cite{Yi-Coder}, CodeLLama 7B \cite{CodeLLama}, MagiCoder 6.7B \cite{wei2023magicoder}, DeepSeekCoder 7B \cite{guo2024deepseek}, CodeGemma 7B \cite{team2024gemma}, QwenCoder-2.5 7B \cite{QwenCoder}, GPT-3.5 \cite{GPT-3.5}, DeepSeek-V3 \cite{DeepSeek}) using the formulated prompts to obtain different outputs for every single task. Those outputs are then used to compute a task score per CodeLLM based on collected metrics from the obtained code snippets. The task scores are then processed using an IRT model to estimate the characteristics of each task (namely, difficulty and discriminant) and the abilities of each model. From there, a more nuanced analysis can take place. We highlight the difference between the two benchmarks under study, showing that \textit{ClassEval} tends to have more extreme tasks in terms of difficulty (easy and difficult) than \textit{HumanEval+}. We further the analysis by extracting from the available tasks, respectively 17 and 21 topics for both \textit{HumanEval+} and \textit{ClassEval}, highlighting some topics of interest (such as tasks dealing with sequences of numbers in \textit{HumanEval+}) that tend to have a higher level of difficulty and be discriminant between CodeLLMs. We also analyse the prevalence of specific program constructs (e.g., variable assignments) by leveraging the Abstract Syntax Tree (AST) of the generated code, uncovering trends related to task characteristics. For example, we observe that the use of conditional statements increases with task difficulty, regardless of code length. Finally, we contrast the difficulty evaluated using \textit{TaskEval} with human annotators and find a difference between both evaluations, emphasising that the difficulty of the tasks for CodeLLMs should not be assessed by humans as is. To guide our study, we formulate the following Research Questions (RQs):

\begin{itemize}
    \item[\textbf{RQ1}] \rqone
    \item[\textbf{RQ2}] \rqtwo
    \item[\textbf{RQ3}] \rqthree
    \item[\textbf{RQ4}] \rqfour
\end{itemize}

In summary, this paper makes the following contributions:

\begin{itemize}
    \item We highlight the limitations of assessing difficulty using traditional evaluation with a single prompt.
    \item We present \textit{TaskEval}, a framework for evaluating the characteristics of code generation tasks for LLMs,    
    \item We perform a cross-analysis of task characteristics, examining their relation to task topics, program constructs (e.g., conditions, variable assignments) used in task codes, and comparisons with human annotators,
    \item We make the data and code used in our study publicly available for researchers to replicate our results \cite{rep-package}.
\end{itemize}

The rest of this paper is organised as follows. In Section \ref{sec:method}, we describe the methodology we followed to propose \textit{TaskEval}. Our experimental setup for running \textit{TaskEval} is detailed in Section \ref{sec:setup}. We report and then analyse the results obtained in Section \ref{sec:results}. A discussion of our results and the insights for the community is presented in Section \ref{sec:discussion}. We conclude the paper in Section \ref{sec:conclusion} after discussing threats to the validity of our findings in Section \ref{sec:threat}.

\section{Motivating examples}\label{sec:motiv}

To motivate \textit{TaskEval}, we will use two examples of tasks. In our study, we make the distinction between \textit{task} and \textit{prompt}: the \textit{task} is the essence of the functionality to be implemented. For instance, \enquote{Quicksort} or \enquote{Fibonacci sequence} are examples of such tasks. A \textit{prompt} is a way to formulate the task so it can be understood by and fed to an LLM. For instance, \enquote{Write a function to calculate the n-th element of the Fibonacci sequence.} or \enquote{Create a function to return the n-th element of the Fibonacci sequence. The Fibonacci sequence is defined as ...} are two instances of \textit{prompts} for the \textit{task} \enquote{Fibonacci sequence}. As such, each \textit{task} can be expressed by many different (even infinite) \textit{prompts}, including different wordings, amount of contextual information about the task, like examples, chain-of-thoughts, etc. To emphasise the importance of using multiple prompts for estimating the task's characteristics, our approach employs multiple prompts, in contrast to the traditional method, which relies solely on a single prompt.

In the first example (see Listing \ref{list:example-2}), extracted from \textit{HumanEval+}, which involves checking if an array is correctly sorted in ascending order without more than one duplicate, most CodeLLMs failed to generate a correct code when using the original prompt. Pass@1 on this prompt is 0 for all models except DeepSeek-V3 with 1.0 and QwenCoder with 0.2. The main source of error is the handling of duplicates. We believe this issue arises because most CodeLLMs struggle with cases involving ``multiple duplicates'', and the provided examples are insufficient. However, the difficulty level computed by our approach is $0.17$ — a relatively low value, among the lowest in this benchmark. Indeed, by slightly modifying the prompt, better-generated code can be obtained compared to the one obtained with the original prompt. For instance, the example we give in Listing \ref{list:example-2} shows that by wording the condition with \enquote{occurs more than twice}, the CodeLLMs manage to generate more correct codes. In that case, the Pass@1 improves drastically, ranging from 0.6 to 1.0 for all models. Note that, in that case, having an example did not appear to have helped the CodeLLMs generate a correct code, contrary to reformulating the prompt.

\begin{lstlisting}[language=python, caption={Example of discrepancy on a Sorting Task. (\textbf{Top}) Original Prompt, (\textbf{Bottom}) One of our generated prompt.}, label={list:example-2}]
def is_sorted(lst):
   """Given a list of numbers, return whether or not they are sorted in ascending order. If list has more than 1 duplicate of the same number, return False. Assume no negative numbers and only integers.

   Examples
   is_sorted([5]) -> True
   ...
   is_sorted([1, 2, 2, 2, 3, 4]) -> False
   """

def is_sorted(lst):
   """ Create a function called 'is_sorted' to determine if a list of non-negative integers is in ascending order. The function should also return False if any integer in the list occurs more than twice.
   """
\end{lstlisting}

In the second example (see Listing \ref{list:example-1}), extracted from \textit{ClassEval}, which involves coding a part of a basic restaurant management system with the method `add\_dish`, we generated two prompts starting from an original one: the first one contains relatively few information (\textit{top}), obtained by having the prompt simply reformulate the original prompt; the second one (\textit{bottom}) contains more information by adding natural description of the inner working of the function, obtained by describing partly the available oracle. Both prompts also included the class constructor and description to provide the CodeLLMs with general context. In the first case, the Pass@1 was 1.0 only for DeepSeek-V3 and QwenCoder, the rest of the models oscillating between 0 and 0.2. Several CodeLLMs struggled on multiple points: the implicit fact that, if the dish is added, the quantity should be lowered, hallucinating undefined methods, handling the non-required sales, etc. However, our approach evaluates the difficulty of the approach to $0.2$, relatively low difficulty. Indeed, if instead a more thorough, yet high-level, description of the intended functionality is given, such as is the case with the \textit{bottom} prompt, the Pass@k improves for all models, reaching between $0.6$ and $1.0$.

\begin{lstlisting}[language=python, caption={Example of discrepancy on a SQL task. For "Function to implement": (\textbf{Top}) Low-level information prompt generated, (\textbf{Bottom}) Higher-level information prompt generated}, label={list:example-1}]

class Order:  
    """
    The class manages restaurant orders by allowing the addition of dishes, calculation of the total cost, and checkout.
    """

    def __init__(self):
        """
        Initialize the order management system
        self.menu stores the dishes of resturant inventory
        menu = [{"dish": dish name, "price": price, "count": count}, ...]
        self.selected_dishes stores the dished selected by customer
        selected_dish = {"dish": dish name, "count": count, price: price}
        self.sales stores the sales of each dish
        sales = {dish name: sales}
        """
        self.menu = []
        self.selected_dishes = []
        self.sales = {}

    ...

    # Function to implement
    def add_dish(self, dish):
       """ Append the dish to 'self.selected_dishes' provided that the dish is present in 'self.menu' and adequate quantities are available. Return 'True' on successful addition, and 'False' otherwise.
       :param dish: dict, the information of dish. dish = {"dish": dish name, "count": count, price: price}
        :return: True if successfully added, or False otherwise.
       """

    def add_dish(self, dish):
       """ Incorporate the dish into 'self.selected_dishes' only if it is present in 'self.menu' with sufficient quantities. Examine each dish in 'self.menu' to see if the name matches the provided dish. If a match is discovered and the current quantity exceeds or meets the needed quantity, update the stock by decreasing the quantity in 'self.menu', then add it to 'self.selected_dishes'. If the addition is successful, return 'True'; otherwise, return 'False'.
       :param dish: dict, the information of dish. dish = {"dish": dish name, "count": count, price: price}
        :return: True if successfully added, or False otherwise.
       """
\end{lstlisting}

Through these examples, we see that traditional assessment may be limited in accurately evaluating the characteristics of tasks, which can, in turn, hinder benchmark analysis. More generally, if we were to compute the pass@1 per task for different rephrasing, while we would observe limited change in the average Pass@1 (0.01 to 0.03 standard deviation for \textit{HumanEval+} across models and rephrasing and 0.02 to 0.05 for \textit{ClassEval}), the Pass@1 on the same task can double from 10\% to 30\% of the benchmarks tasks for a given model, that is the correctness can be drastically altered by the prompt formulation. We observe similar results if we consider the level of information. In those cases, it is then not possible to infer the \enquote{difficulty} of the task; at best, the difficulty of a given prompt formulation could be deduced. Confronted with this issue, our approach aims to account for this disparity by giving a difficulty score that accounts for possible prompt variation as well as disparity between models.

\section{TaskEval}\label{sec:method}
In this section, we discuss the methodology followed to propose \textit{TaskEval}. After an overview of the framework, we describe how we generate different prompts for programming tasks. We then delve into the details of leveraging CodeLLMs to generate code for our prompts, fitting the IRT model and measuring the characteristics of tasks for CodeLLMs.

\subsection{General Overview of \textit{TaskEval}}
\textit{TaskEval} uses existing benchmarks to operate. Such benchmarks are generally composed of an ensemble of \textit{tasks} which are represented by a single \textit{prompt} per task. \textit{TaskEval} aims to evaluate the characteristics of those \textit{tasks}, and not of the prompts, for CodeLLMs to probe benchmarks and analyze their behavior regarding those tasks (too difficult? too easy? etc.)

\begin{figure*}
    \centering
\includegraphics[width=\textwidth]{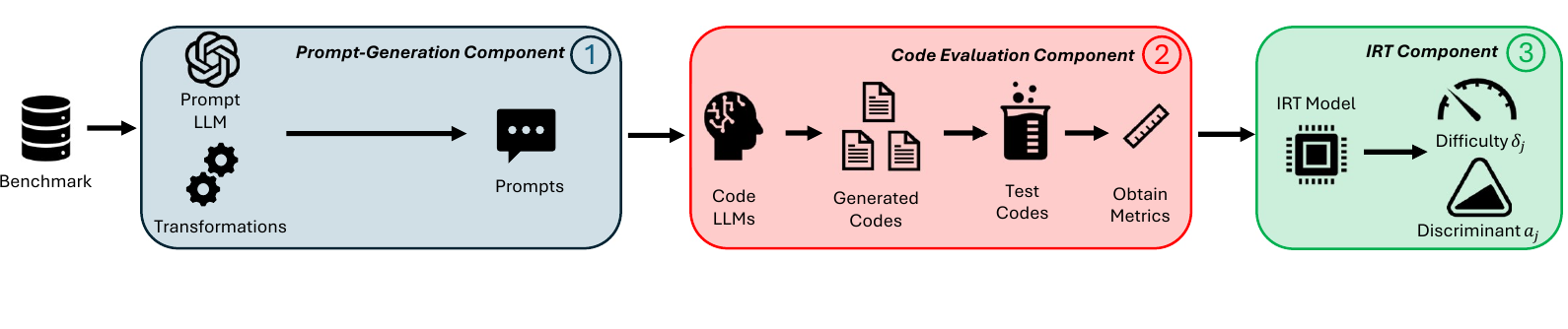}
\caption{Overview of \textit{TaskEval} Framework.}
\label{fig:gen_framework}
\Description[TaskEval framework process]{The figure details the three step process of TaskEval. Process involves: 1) Generating the promtps based on transformations, 2) Query the CodeLLMs with the generated prompts and verify the generated code correctness, 3) Fit the Item Response Theory model on the data}
\end{figure*}

\begin{algorithm}
\caption{TaskEval high-level algorithm}\label{alg:method}
\DontPrintSemicolon
\KwData{Benchmark under-test $D$ with $n$ tasks; Set of prompts transformations $\{t_1, ..., t_T\}$; Prompt LLM $Prompt_{LLM}$; Set of Code LLMs $\{CodeLLM_1, ..., CodeLLM_m\}$; Number of temperature generation $S$}
\KwResult{IRT model parameters $\delta$, $\alpha$, $\theta$}
\emph{Prompt Generation}\;
Initialize \textit{promptsArray} an empty 2D array of size $n \times T$\;
\For{i $\leftarrow$ 1 \KwTo $n$}{
    \textit{basePrompt} $\leftarrow$ $D_i$\;
    \For{j $\leftarrow$ 1 \KwTo $T$}{
        \textit{basePrompt} $\leftarrow$ $Prompt_{LLM}$(\textit{basePrompt}, $t_j$)\;
        \textit{promptsArray}[i][j] $\leftarrow$ \textit{basePrompt}\;
    }    
}

\tcc{Manual verification occurs here}

\emph{Code Generation}\;
Initialize \textit{scoreArray}, a zero-filled 2D array of size $m \times n$\;
\For{i $\leftarrow$ 1 \KwTo $m$}{
    \For{j $\leftarrow$ 1 \KwTo $n$}{
        \For{t $\leftarrow$ 1 \KwTo $T$}{
            \For{s $\leftarrow$ 1 \KwTo $S$}{
                \textit{codeGenerated} $\leftarrow$ $CodeLLM_i$(\textit{promptsArray}[j][t], s)\;
                \tcc{Using test oracles from benchmark $D$}
                \textit{scoreArray}[i][j] = \textit{scoreArray}[i][j] + isCodeCorrect(\textit{codeGenerated})\; 
            }
        }
        \textit{scoreArray}[i][j] = \textit{scoreArray}[i][j] / ($T \times S$)\;
    }
}

\emph{IRT Compute}\;
\tcc{Computing for $n$ items and $m$ respondents}
$\delta$, $\alpha$, $\theta$ $\leftarrow$ IRT(\textit{scoreArray})\; 

\end{algorithm}

A general overview of \textit{TaskEval} is given in Figure \ref{fig:gen_framework} and on the pseudo-algorithm \ref{alg:method}. \textit{TaskEval} is operated in three steps: \tikz[baseline=(char.base)]{
  \node[shape=circle, draw=blue, inner sep=1pt, text=blue] (char) {1};
} First, transformations on the prompts are identified. Those transformations are any dimension that can lead to a variety of prompts for an LLM while keeping the essence of the task. In a way, such transformations can be seen as some metamorphic transformations \cite{Metamorphic} as they should not alter the semantics of the task. In our study, we focus on two transformations: \textit{Rephrasing} and \textit{Context Information}. Those transformations are then used in conjunction with each task to have a \textit{Prompt LLM} generate prompts following the transformations' specifications. \tikz[baseline=(char.base)]{
  \node[shape=circle, draw=red, inner sep=1pt, text=red] (char) {2};
} Secondly, the obtained prompts are fed into multiple CodeLLMs to generate code. The goal is to evaluate how each CodeLLM performs when confronted with different prompts. Using multiple CodeLLMs allows us to capture the diversity of architectures and training settings, which may respond differently to a given prompt. Generated codes are evaluated on available tests, and their results are aggregated into a score per task and per CodeLLM. Using these scores, an Item Response Theory~\cite{baker2001basics} (IRT) model is trained \tikz[baseline=(char.base)]{
  \node[shape=circle, draw=codegreen, inner sep=1pt, text=codegreen] (char) {3};
}. The model estimates the latent attributes of items (tasks) and respondents (CodeLLMs) based on the observed responses of all tasks and all CodeLLMs. In the following, we describe in more detail the different parts of our approach.

\subsection{Generating Different Prompts based on Transformations}\label{sec:step1}

The first step involves choosing transformations to generate different prompts for the task. We want those transformations to introduce variabilities in our prompts to represent the many ways a CodeLLM could be prompted for the same task. However, those transformations should not introduce changes that alter the semantics of the tasks. In our case, we use two types of such transformations, but any number could be used and combined. 

First, we use the \textit{Context Information} transformation. Indeed, a straightforward way to act on the variability of a prompt is to simply disclose or withdraw certain information in the prompt, while not changing the task. EvoEval \cite{EvoEval}, for instance, also proposes to act on the difficulty. However, to modify the difficulty, they do so by adding/removing constraints in the prompt, thus modifying the semantics of the task, which is not desirable in our case. In our study, we instead propose to divide the information provided in the prompt into three levels. Those three levels of prompts are generated incrementally using the previous level as the starting point. We assume that prompts with more information should be more likely to lead to correct code for CodeLLMs \cite{deng2023rephrase, EvoEval}. The levels are defined as: 

\begin{itemize}
    \item \textit{First level}: contains a minimal amount of information that is the inputs/outputs of the task, as well as a high-level description of what is intended. We could use the original prompt in benchmarks for each task as the first level, yet Siddiq et al. \cite{PromptQuality} showed that there can be inconsistencies across tasks' prompts in benchmarks. Thus, we prefer to generate a new prompt to have similar formatting across all tasks and levels to reduce biases. This level is the closest one to the original prompt in benchmarks and contains the same level of information regarding the tasks.
    \item \textit{Second level}: further adds in the prompt a description of the targeted function in natural language using available oracle code. However, it does not include any direct reference to the function (variable names, helper functions used, ...) unless it was explicitly mentioned in the original prompt of the benchmark. 
    \item \textit{Third level}: complete the prompt with direct references to the function, such as variable names, and helper functions used similarly to a pseudo-algorithm but in natural language.
\end{itemize}

We chose to devise these three levels based on our observations. They prove to generate a sufficient amount of fine-grainedness and to discriminate between different levels of information. Trying to fit an additional level (e.g, between level 1 and 2) did not lead to noticeable changes. We present an example of different prompt levels obtained for a task in Listing \ref{list:levels}.

\begin{lstlisting}[language=python, caption={Example of different prompt levels for the task \textit{unique} from HumanEval+, with the oracle code at the bottom.}, label={list:levels}]
def unique(l: list):
    Level1: """ Write a function named 'unique' that returns a sorted list of unique elements from a given list. """
    Level2: """ Write a function named 'unique' which takes a list as input and aims to return a sorted list containing only the unique elements from the input list. The function should first convert the list to a set to remove any duplicates and then convert it back to a list which should be sorted before returning. """
    Level3: """ Write a function named 'unique' which takes an input list 'l' and returns a list containing only the unique elements from 'l', sorted in ascending order. The function should first convert 'l' into a set, using 'set(l)', to remove any duplicates, then convert this set back to a list, and finally return this list after sorting it in ascending order using 'sorted()'. """
    return sorted(set(l))
\end{lstlisting}

The second transformation used is \textit{Rephrasing}, which is paraphrasing a given prompt. Indeed, how the prompts or instructions are formulated has been shown to have an impact on LLM responses \cite{MultiPrompts, InstructionDiff, ReCode}. Thus, similarly to these studies, we try to see if different ways of rephrasing a given prompt for a CodeLLM can affect how the CodeLLM addresses the task. To rephrase our prompts, we will use a similar approach as done by Gonen et al. \cite{RephrasePaper}, that is, prompting an LLM to rephrase a given text. More precisely, we prompt the PromptLLM with a general meta-prompt rephrasing instruction “You are given a Python function in \#\#\# Function and a prompt in \#\#\# Prompt. Your task will be to reword the given prompt in order to generate a variation of said prompt.”, alongside the actual oracle code of the task (in \#\#\# Function) and the prompt to be rephrased (in \#\#\# Prompt). The LLM is instructed to generate all rephrasings one after another, using one prompt. The full prompt used can be found in our replication package [8]. We give an example in Listing \ref{list:rephrase}. Other techniques could be used to assist in rephrasing/information, such as Chains-Of-Thoughts \cite{CoT}, but assessing the effectiveness of additional techniques is out of the scope of this study.

\begin{lstlisting}[language=python, caption={Example of different prompt levels for the task \textit{unique} from HumanEval+, with the oracle code at the bottom.}, label={list:rephrase}]
def unique(l: list):
    Base: """ Write a function named 'unique' that returns a sorted list of unique elements from a given list. """
    Rephrasing 1: """ Develop a function titled 'unique' which outputs a list sorted with distinct elements extracted from a provided list."""
    Rephrasing 2: """ Construct a function called 'unique' that delivers a sorted list containing only unique elements from an input list."""
    Rephrasing 3: """ Create a function known as 'unique' that yields a list, sorted and containing only unique items from a specified list."""
    Rephrasing 4: """ Design the function 'unique' to produce a list where elements are unique and sorted, derived from a given list."""
    Rephrasing 5: """ Formulate a function named 'unique' that generates a sorted list of distinct elements originating from an input list."""
    return sorted(set(l))
\end{lstlisting}

Using those two transformations, we come up with general templates that can be used to generate prompts (see replication package \cite{rep-package}). Such general templates are then tailored for each task in the benchmark under consideration. Those templates are fed to a \textit{Prompt LLM} (see Figure \ref{fig:gen_framework}) to generate different prompts for a given task. The way prompts are generated is as follows: we start from the original docstring to have a common starting point for all prompts. Then, we apply the template to obtain different prompt levels with \textit{Context Information} for a given task. Then, we apply the obtained prompts to the \textit{Rephrasing} template. In our case, we first remove the examples from the original prompt, if any, before using our templates with the \textit{Prompt LLM}. While removing examples can decrease the performance of CodeLLMs \cite{CodeMind}, this allows for better control over the level of information we inject into our prompts. Indeed, examples can vary from simple, basic examples to showcasing corner cases of the task. As such, this might artificially bias the output of CodeLLMs and the potential difficulty of the task because of the quality of the examples. 

At each step of our generation, that is, after using each template on the prompts, we do a sanity check by manually checking them. This is accomplished by the first author and cross-checked by the second author independently. We make sure that the \textit{Context Information} levels are respected and the \textit{Rephrasing} does not alter the semantics of the tasks. This is done in order not to have tasks that are artificially considered hard further down the line because of misleading prompts for the CodeLLM and not because of the task itself.

\subsection{Collecting metrics for code snippets}\label{sec:met_task}

The second step is to gather metrics for a given task. To do so, we leverage several CodeLLMs that will serve as our evaluators for code generation. For each prompt per task, each CodeLLM is asked to generate code with varying seeds using the sampling mode of CodeLLMs. By seed, we mean a random number used in the sampling-based generation mode of CodeLLMs. The rationale is that most CodeLLMs, especially non-open-source ones, are used with this non-greedy approach to generate code. Moreover, using the sampling method for the generation allows us to have a wider variety of generated code in order to evaluate the difficulty of each task. Indeed, we assume an easier task should not be as impacted by the stochasticity of the generation compared to a harder task. The obtained code is then processed and executed. Sample codes were then executed against available test inputs in the benchmark.

Once we have obtained all the codes for all prompts of all tasks and executed them against tests, we can collect different metrics. In our case, we are mainly interested in the \textit{Functional} correctness of a code sample. \textit{Functional} correctness quantifies whether a given code sample is correct or not using available test cases. Formally, the \textit{Functional} correctness $Func$ for a code sample $s$ generated by CodeLLM $LLM$:

\begin{equation}
    Func_{LLM}(s) = \left\{
    \begin{array}{ll}
        1 & \mbox{if } \forall (i, o) \in T, s(i) = o \\
        0 & \mbox{else }
    \end{array}
    \right.
\end{equation}
where $T$ is the set of test cases symbolized by an input $i$ and an output $o$. That is, the code sample passes all test cases, i.e., gets a value of $1$. Intuitively, we expect a harder task to lead to fewer prompts producing correct codes across rephrasing and different levels of context information.

The previous score is obtained for a single code sample (i.e., using a single prompt and with a single seed). However, we want to obtain a score at the \textit{task} level. Thus, we need to aggregate all scores for each prompt/seed of the same task. In the following, we note $j$ for the task index, $c_k$ for the $k^{th}$ information context, $r_l$ for the $l^{th}$ rephrasing, and $p$ for the seeded generation. As such, for example, $s_{1, c_{1}, r_{1}, 1}$ is the code sample obtained for the first task, using the low-level information context with the first rephrasing with the first seed generated. As such, the aggregation over all variables for a task is as follows:

\begin{equation}
    \label{eq:score}
    score_{LLM}(s_{j}) = \frac{1}{N_j}\sum_{c_k} \sum_{r_l} \sum_p Func_{LLM}(s_{j, c_k, r_l, p})
\end{equation}

where $N_j$ is the total number of prompts for the task $j$. The final score $score_{LLM}(s_j)$ represents the score for the task $j$ using a given \textit{LLM}. We will use the scores for all CodeLLMs to determine the task's parameters $j$.

\subsection{Item Response Theory modelling}

Since we have a metric measured across all tasks and multiple CodeLLMs, we can use these values as observed responses in an IRT model. IRT models, commonly used for analysing and scoring psychological tests \cite{zanon2016application}, have already been used in NLP settings \cite{rodriguez2021evaluation, lalor2016building}. In the literature, several IRT models exist depending on the observed response distribution or the number of parameters of the model. A widely used model is the 2-Parameter (2PL) IRT model. Consider $j = 1, ..., m$ items being assessed by $i = 1, ..., n$ respondents, for example, students answering questions in an exam, leading to true or false outcome $x_{ij} = \{0, 1\}$, thus following a Bernoulli distribution of parameter $p_{ij}$. The 2PL IRT model in this case is defined as:

\begin{equation}\label{eq:bin_model}
    \mathop{\mathbb{E}}[p_{ij} | \theta_i, \delta_j, a_j] = \frac{1}{1 + e^{-a_j(\theta_i - \delta_j)}}
\end{equation}
where $\theta_i$ is the ability of respondent $i$, $\delta_j$ is the difficulty, and $a_j$ is discriminant of item $j$. The IRT model maps a respondent's ability to an expected response using the observed response through a logistic function with a difficulty parameter $\delta_j$ that indicates the "location" on the difficulty range, and a discriminant parameter $a_j$, discriminating items based on their difficulty and the ability of the respondents, reflected in the steepness of the slope of the item characteristic curves. In our case, we do not have a binary response but a continuous one, as the aggregated score at the task level has a value in [0,1]. As such, we will rely on the $\beta^3$-IRT model proposed by Chen et al. \cite{chen2019beta}, which models the observed response as a continuous value in $(0, 1)$ using a Beta distribution. The expected response for the model is defined as:

\begin{equation}
    \label{eq:model}
    \mathop{\mathbb{E}}[p_{ij} | \theta_i, \delta_j, a_j] = \hat{p}_{ij} = \frac{1}{1 + \frac{\delta_j}{1 - \delta_j}^{a_j} \times \frac{\theta_j}{1 - \theta_j}^{-a_j}}
\end{equation}
where $p_{ij}$ is the response of respondent $i$ for item $j$ defined as $p_{ij} \sim  B(\frac{\theta_i}{\delta_j}^{a_j}, \frac{1 - \theta_i}{1 - \delta_j}^{a_j})$, $\theta_i\sim Beta(1, 1)$, $\delta_j \sim Beta(1, 1)$ and $a_j \sim Normal(1, 1)$. In our approach, a respondent will be one of our CodeLLMs, an item a task from a benchmark, and $p_{ij}$ is the score $score_{LLM}(s_j)$ on this task by the given CodeLLM as calculated in Equation~\ref{eq:score}. The parameters are estimated based on the observed response of CodeLLM (that is, the score) using the maximum likelihood estimate. 

\begin{figure}\centering
\includegraphics[width=0.6\textwidth]{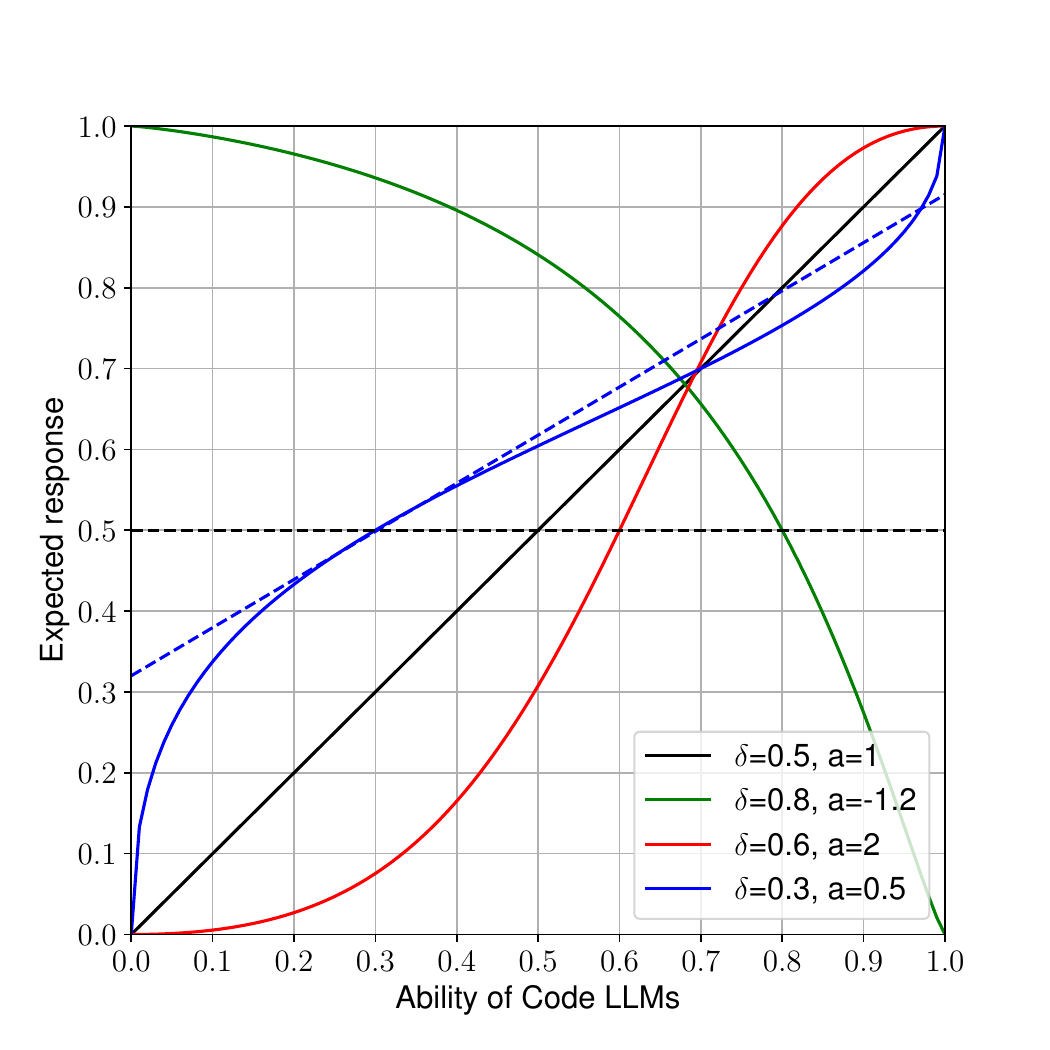}
\caption{ICC between ability $\theta_i$ of a CodeLLM and the expected response $\mathop{\mathbb{E}}[p_{ij} | \theta_i, \delta_j, a_j]$ for four tasks with different $\delta_j$ and $a_j$ values.}
\label{fig:ex_irt}
\Description[Example of Item
Characteristic Curves]{Four curves are shown symbolizing different tasks with varying difficulties and discriminants. This showcases how difficulty and discriminant can be interpreted graphically and how to compare different tasks.}
\end{figure}

To illustrate the response obtainable given the different parameters, in Figure \ref{fig:ex_irt}, we present a theoretical Item Characteristic Curve (ICC) mapping the ability of a CodeLLM with ability $\theta_i$ to the expected response $\mathop{\mathbb{E}}[p_{ij} | \theta_i, \delta_j, a_j]$ following the Equation \ref{eq:model} for four different couple of parameters $(\delta_j, a_j)$  (in plain traits in Figure \ref{fig:ex_irt}). The values are chosen to illustrate the effect of each parameter, as we detail in the following. It should be noted that since we are using functional correctness as our metric, the expected response translates to the likelihood of generating a correct code for a given task and CodeLLM, which is easily interpretable.

From the different curves of Figure \ref{fig:ex_irt}, we can deduce the parameters of the different hypothetical tasks. Indeed, mathematically, the location parameter $\delta_j$ (the difficulty) corresponds to the ability value $\theta_i$ for which the expected response is $0.5$ (black dashed line in Figure \ref{fig:ex_irt}) while the discriminant $a_j$ is such that the slope of the curve is $a_j/(4 \times \delta_j \times (1 - \delta_j))$ at that point (for example, the blue dashed line for the blue item). This explains why the blue task has a difficulty of $0.3$ and a discriminant of $0.5$.

For a task $j$, $\delta_j$ and $a_j$ will ultimately affect how likely a given CodeLLM is to answer this task according to the IRT model: for instance, still for the blue task, a CodeLLM of ability $0.7$ has an expected response of $0.7$. For tasks that have the same $a_j$ but different $\delta_j$, the expected response is always higher for a CodeLLM with a higher ability $\theta_i$: for instance, the expected response on the green task is always higher compared to the black task, no matter the CodeLLM ability, as the difficulty of the green task is lower. In that sense, the difficulty is enough to characterise a task. This is particularly used in the 1PL IRT model (i.e., Equation \ref{eq:model} with $\forall j, a_j = 1$). However, while it is easier to analyse, it is not necessarily the most realistic modelling. For instance, some tasks can have a disproportionately higher/lower expected response for different CodeLLMs of different capacities. This can arise, for instance, if these CodeLLMs were less trained on data similar to this particular task, and are more sensitive to how the task could be phrased. To account for this fact, the 2PL model offers a discriminant coefficient $a_j$ for each task.

When the discriminant starts changing, a higher difficulty equaling a lower expected response no matter the CodeLLM is not always true: still in Figure \ref{fig:ex_irt}, despite the black task being \enquote{easier} than the red one (difficulty of $0.5$ compared to $0.6$), CodeLLMs with an ability higher than $0.6$ will have a higher expected response on the red task. Effectively, the red task is more discriminant than the black task, as it better separates CodeLLMs according to their capacity. As $a_j \in \mathbb{R}$, it is even possible for $a_j < 0$ (green curve in Figure \ref{fig:ex_irt}): in that case, a task can lead to a higher expected answer for a CodeLLM with a lower ability even with a higher difficulty. That last case can be indicative of potential noise \cite{chen2019beta}, annotation errors \cite{rodriguez2021evaluation} or could point towards actual tasks that less capable CodeLLMs have an easier time on for several possible reasons: more facilities with the given input prompts, non-determinism in the generation, example that was memorized or that is similar to what the model was trained on etc.

\section{Experimental setup}\label{sec:setup}

\subsection{Hyperparameters used}
To have different prompts, we choose to generate 3 levels and 6 rephrasings per level. We chose those parameters as a trade-off between generating diverse enough prompts while limiting the number to generate. For a benchmark of $n$ tasks, this effectively means we have to generate $18 \times n$ prompts. We further ask each \textit{CodeLLM} to generate 5 code samples per individual prompt, resulting in $90 \times n$ code samples for each \textit{CodeLLM} to generate. As an example, for a benchmark containing 200 tasks, this process yields \textbf{18,000} generated code for our evaluation \textit{\textbf{per} CodeLLM}. We make available all the prompts and code generated in our replication package \cite{rep-package}.

\subsection{Benchmark used}

In our experiment, we chose two benchmarks: HumanEval+ \cite{HumanEvalplus} and ClassEval \cite{ClassEval}. Those benchmarks were chosen as they cover different types of tasks and dependency levels. HumanEval+ is widely used as a standard benchmark in code generation \cite{BigCode}. It is an extended variant of HumanEval \cite{HumanEval} that contains more tests to assess the correctness of code samples. The benchmark consists of 164 handwritten Python programming problems with a function signature, docstring, body and several unit tests. HumanEval+ tasks are at the function level dependency as they do not require external libraries, besides the Python standard library, or class/file contexts to operate. For the original prompt, we use the docstring as given in HumanEval+. 

ClassEval is a benchmark of 100 handcrafted classes coded in Python. Each class contains a description and, on average, 4 methods each with a docstring. We consider a task to be one of the 400 methods of ClassEval. Contrary to HumanEval+, ClassEval tasks are more similar to practical code used in projects and rely on a class context to operate. ClassEval proposes three prompting strategies: holistic, incremental and compositional. In our case, we use the compositional strategy, that is, providing the class context to the LLM and asking it to complete one method of the class, only modifying the docstring of the method to complete through our crafted prompts. We do not make use of the other generation approaches as it reduces the control we have over the prompts and could introduce biases: the holistic approach, as it implies asking the LLM to generate the whole class from scratch, can impact the performance artificially. Similarly, the incremental approach requires the reuse of codes generated by the LLM for other methods when generating new ones, which could also alter the difficulty. As we need to generate a high number of code samples and prompts, we reduce the number of tasks to evaluate by doing a stratified sampling at the 95\% confidence interval. For 400 methods, this gives us 197 methods to sample, which we round up to 200. Those 200 methods are sampled equally across classes, resulting in 2 methods per class available. Methods chosen are available in our replication package \cite{rep-package}.

\subsection{Models used}

In our experiments, we use GPT-4 \cite{GPT-4} as the \textit{Prompt LLM} (to generate the different prompts). We do so as GPT-4 is the state-of-the-art LLM\footnote{at the time the experiments were started} and so should provide answers with the best quality, which would require a minimum of human supervision. For the \textit{CodeLLM}, we use 8 LLMs with different architectures: YiCoder-1.5B \cite{Yi-Coder}, CodeLLama 7B \cite{CodeLLama}, MagiCoder 6.7B \cite{wei2023magicoder}, DeepSeekCoder 7B \cite{guo2024deepseek}, CodeGemma 7B \cite{team2024gemma}, QwenCoder-2.5 7B \cite{QwenCoder}, GPT-3.5 \cite{GPT-3.5}, DeepSeek-V3 \cite{DeepSeek}. We have not used GPT-4 as \textit{CodeLLM} to avoid biases, as it is used to generate the prompts fed to the \textit{CodeLLM}. For all models, we used their instructions-tuned versions so that the models can handle the different prompts properly. This gives us a diverse array of architecture and training procedures. For all code generation in \textit{TaskEval}, since we are using sampling generation, we set the temperature hyperparameter to 0.8 as it is one of the temperature settings commonly used in other similar studies \cite{ExploringHallu, HumanEval, HumanEvalplus, CodeLLama}. In all cases except GPT models (for which we use OpenAI API) and DeepSeek-V3 (for which we use DeepSeek API), we use the HuggingFace \cite{HFmodels} implementation of the models to allow for replication. Ranking of the LLMs' performance on our prompts follows the assessment obtained on the original benchmarks.

\subsection{IRT model fit}
For the IRT model, we use the library Birt-GD \cite{Beta4} which contains a gradient descent-based Maximum Likelihood Estimation of the $\beta^3$-IRT algorithm. We use the implementation exactly as is, except we modify the initial values of the parameters: $\delta_j$ and $\theta_i$ are initialized as the average of scores for each task/CodeLLMs while $a_j$ is initialized as the Pearson correlation coefficient between the $\theta_i$ and the scores on each task of the $i^{th}$ CodeLLM. This last initialisation helps to solve the symmetry issue of the $\beta^3$-IRT model \cite{Beta4}. The implementation details can be found in our replication package \cite{rep-package}. We use the scores collected as described in Section \ref{eq:score}. Then, we will fit two IRT models, one for \textit{HumanEval} and one for \textit{ClassEval} data. The $R^2$ \cite{Beta4} after fitting the IRT models to our data are $0.87$ and $0.90$, respectively, showing an accurate fit.

\subsection{Research Questions}\label{sec:rqs}

In this part, we describe our RQs and how we plan to address them.

\textbf{RQ1: \rqone} In our first RQ, we will analyse the tasks of both benchmarks through the prism of our IRT model. To do so, we will first compare the two benchmarks based on the expected probability of response of each CodeLLM, that is $E[p_{ij} | \theta_i, \delta_j, a_j]$, through their cumulative distributions across tasks. Besides observations of the cumulative distributions, we apply the Anderson-Darling (AD) \cite{scholz1987k} test\footnote{The Anderson-Darling (AD) test is a statistical test to determine whether two independent samples come from the same (but unspecified) continuous distribution. The result of the test is a “p-value” which represents the probability of observing differences among the sample distributions as extreme as those measured, assuming both samples come from the same underlying distribution. A lower p-value suggests stronger evidence that the samples come from a different distribution. Generally, a threshold of 0.05 (95\%) is selected.}, a statistical test to assess the difference between the two empirical distributions, between the distributions across benchmarks for the same CodeLLM. Then, using the obtained discriminant $a_j$ and difficulty $\delta_j$, as calculated by the IRT model, we will draw a portrait, task-wise, of both benchmarks. This will allow us to describe tasks present in the benchmark qualitatively.

\textbf{RQ2: \rqtwo} In this RQ, we explore how tasks' topics can affect how the difficulty/discriminant parameters are evaluated. We chose to explore at the topic level as it gives a good trade-off between a more nuanced analysis than the benchmark level while allowing us, compared to individual tasks analysis, to draw some conclusions over a group of tasks. To obtain meaningful topics, we will leverage the topic modelling approach with BERTopic \cite{BERTopic}, as was used in similar studies when dealing with natural language data \cite{ChatGPTHCI, ChatGPTEdu, ChatGPTTrust}. BERTopic is a topic modelling technique that uses transformer-based embeddings (like BERT) to capture the semantic meaning of documents. It first converts documents into dense vectors, then reduces their dimensionality using UMAP \cite{mcinnes2018umap} to make clustering feasible. These vectors are clustered using HDBSCAN \cite{mcinnes2017hdbscan}, which groups similar documents without requiring the number of topics in advance. Each cluster’s most representative words are extracted using a class-based TF-IDF (c-TF-IDF) method to form interpretable topics. The result is a set of meaningful topics with associated keywords and document groupings.

For each benchmark, the topic modelling clusters all tasks using the prompts of Level 1 before rephrasing. We use this specific prompt type to have similar formatting across tasks while being as close as possible to the original prompt, as mentioned in Section \ref{sec:step1}. We consider a topic to be valid only if it contains at least three tasks. This criterion ensures that in \textit{ClassEval}, we avoid ending up with an excessive number of topics, each containing only two tasks, for example, which could occur due to the sampling methodology used to select tasks from the benchmark. As the process will inherently label some tasks as ``noise", noisy tasks are not considered in the analysis. This results in \textbf{17} topics for \textit{HumanEval+} and \textbf{21} topics for \textit{ClassEval}. Those topics constitute tasks that have a similar functionality/goal. For the topics' names, we used a refined version of the names provided by BERTopic. To refine the name of the topic, the first two authors manually check the tasks assigned in each topic by BERTopic, notably the prompt given as a description of the tasks. Based on this information, the name was updated by the authors independently before meeting to agree on a common name. The goal of this step was to give a topic name both more explicit than what BERTopic will assign as well as more general. For instance, one of ClassEval's original topic names is \enquote{sql\_select\_female\_under\_age\_table\_name\_query}. This is a BERTopic aggregation of keywords identified in functions’ descriptions dealing with SQL. To make it more understandable, and since no other topic focused on SQL requests, it was decided after independent analysis and a discussion to modify the name to “SQL Requests”, which is more understandable. Then, for each topic, the mean accuracy of each CodeLLM as well as the difficulty and discriminant among all tasks in the topic is calculated and analysed.

\textbf{RQ3: \rqthree} In this RQ, we aim to explore the relationship between the difficulty or discriminative power of a task for a CodeLLM and the program constructs (e.g., If statements, For-loops, Function Calls, etc.) present in the generated code snippets used to solve the task. To achieve this, we will analyse the Abstract Syntax Tree (AST) of each code fragment generated for the task. Selecting only correct code samples (i.e. the ones passing all the tests) would be limiting, as code samples can be wrong to different degrees depending on the bugs affecting the code sample \cite{tambon2024bugs}. It could also bias the analysis towards code fragments obtained for higher-level prompts that are more likely to generate correct code. However, we also should not take any code fragment, as we may end up with generated code not even try to solve the task if the CodeLLM misunderstood the prompt. To strike a balance, we will only consider code fragments that are similar enough to a correct one. We do so by using CodeBLEU \cite{CodeBLEU}, which measures the syntactical correctness of code samples. We thus measure the similarity between each code fragment and a correct code fragment that passed the tests and only retain the code fragments with a similarity greater than a pre-defined threshold. Note that there is always one correct code fragment per task as a reference, which is provided in the benchmark. We chose CodeBLEU as it is a widely used metric in code generation tasks \cite{CodeXGlue, RobustnessCodeGen, StarCoder, CodeT5}. Once the code fragments that are similar enough have been obtained, we can proceed to collect the AST node type. We used the ast package from the Python standard library. For each node extracted in that way, we register the node type as per the AST. Once the nodes are extracted, as the number of lines of code could bias the results (e.g., harder tasks may require more lines of code to be solved), we normalise by the number of lines of code of each fragment. Finally, we calculate the average number of each program construct across all code fragments within the same task. To prevent averaging over too few code fragments, we only retain the tasks for which all models produced at least 10\% of similar enough code fragments. Similarly, to prevent calculating correlations for node types that are barely represented across tasks, we ensure that at least 10\% of the tasks include a particular node type to include it for comparison. This gives us the number of each node type, across tasks and CodeLLMs, which will be compared with the difficulty/discriminant of each task using the Kendall-$\tau$ \cite{kendall1938new}\footnote{Kendall's tau test is a non-parametric test that assesses the strength and direction of association between two ranked (ordinal) variables. The tau value measures the correlation between the rankings, ranging from –1 (perfect inverse agreement) to +1 (perfect agreement), with 0 indicating no association. A p-value can be calculated as done in the AD test. In that case, a lower p-value provides stronger evidence that there is a statistically significant association between the two variables.} to measure the correlation while handling ties. The node types with a significantly positive correlation will signify an increasing trend with difficulty. Furthermore, we will also use the AD test to check whether the distributions of the top/bottom 50\% difficult/discriminant tasks exhibit different behaviour for each programming construct and each CodeLLM.

The threshold to use in CodeBLEU is a critical point, as it can drastically affect the outcome. Indeed, a higher threshold will mean that code fragments are likely very close to a correct solution, but their number will be low, which might limit any test relevance. On the other hand, a low threshold means a sufficient number of code fragments per task, but an increased possibility of including code fragments not implementing the desired functionality. As a manual analysis to establish a reliable threshold when dealing with such a large number of samples is impractical, we will instead choose to cross-compare using multiple thresholds. That is, we will retain code fragments with a similarity above {0.4, 0.5, 0.6, 0.7, 0.8}, respectively. Bounds were chosen empirically: 0.4 is the threshold for which CodeBLEU authors established a correlation with an average human judgment of 4 out of 5, and a similarity above 0.8 (e.g. 0.9) would return too few samples. Then, when calculating the Kendall-$\tau$ and the AD test for each programming construct, we will only retain those programming constructs for which there is a statistical significance across all thresholds at the 0.05 threshold. We employ the Holm-Bonferroni \cite{holm1979simple} correction to ensure that a significant p-value is observed across all thresholds, that is, we iteratively verify $p_i < 0.05 / (5 - (i - 1))$ for i in [1; 5] and $p_i$ is the most significant p-value at iteration i. Doing so, we ensure that the observed statistical significance is robust to the choice of the threshold. In our experiment, this process resulted in between 80\% and 94\% for ClassEval and between 88\% and 99\% for HumanEval+ of tasks being considered in our analysis, depending on the threshold used. Moreover, while very high difficulty tasks (>0.9) were generally removed because of too low a number of valid fragments, we ensured that at least 15\% of tasks with a difficulty of 0.5 or more were present, no matter the CodeBLEU threshold used.

\textbf{RQ4: \rqfour} In our final RQ, we aim to compare task difficulty assessments made by human annotators with those generated by our approach. To achieve this, we collect human-labelled difficulty ratings for both \textit{HumanEval+} and \textit{ClassEval} by employing paid developers as crowd workers on the Prolific platform\footnote{\url{https://www.prolific.com/}}, which was used in previous studies \cite{zid2024study}. To have the crowd-workers estimate the difficulty, the most straightforward approach would be to have them implement the selected tasks. However, a pilot study with 10 developers for 12 tasks we conducted revealed that several participants are inclined to use LLMs to code for themselves, despite it being forbidden as advertised in the study. Given our limited control over how recruited participants might use LLMs, we adopted an alternative experimental setup that leverages effort estimation techniques, such as Planning Poker \cite{Grenning02}, which is familiar to developers. Since our study involves a crowd of workers, we followed the experimental design proposed by Alhamed et al. \cite{Alhamed21}, which focuses on effort estimation when relying on a crowd of non-experts. Their approach simulates real-world effort estimation processes but eliminates the need for developers to resolve estimation conflicts. The method is structured into iterative rounds. In the initial round, individual participants are asked to estimate the difficulty of a task through a time-based scale (e.g., one hour or one day) using solely the information included in the task description. Participants are also required to provide comments justifying their estimates. Once a sufficient number of responses are collected, the level of agreement among participants is calculated. If the agreement exceeds a predefined threshold, the task's effort estimation is determined by taking the median of the responses. If the threshold is not met, a subsequent round is conducted with a new set of participants. These participants are provided with the task information, along with the comments and results from the previous round. This process repeats until either an agreement is reached or a maximum number of rounds is completed.

In this study, we implemented Alhamed et al.'s approach with $5$ participants per task for a given round and a threshold of agreement level 
set to $0.6$ (i.e., Substantial Agreement). Given that we have a limited budget (as crowd-workers are paid based on the time spent), we aimed for $50$ developers in the first round. This results in $120$ tasks (out of the $364$ of both benchmarks) being considered, which were sampled at random. Tasks were then sorted into $10$ questionnaires of $12$ assessments ($6$ for \textit{HumanEval+} and $6$ for \textit{ClassEval}). To balance the time required to complete the questionnaires, we categorised tasks into three groups based on the number of lines of code and the cyclomatic complexity of their oracles before sampling. All questionnaires can be found in our replication package \cite{rep-package}. Each assessment is decomposed into several parts: First, several prompts for the same tasks are provided to the participants, who have to summarise what the task is about. This allows us to check that the participant properly read the task. Prompts given to the participants, for each task, were sampled from the prompts CodeLLMs were evaluated on. Thus, both humans and LLMs had access to a similar description of the task. We made sure to always include the original prompt in the sampled prompts given to participants. Then, the participant is invited to estimate the duration it would take them (from 1 minute to over 40 minutes) as well as the subjective difficulty of the task (using a Likert scale from "Very Easy" to "Very Difficult"). They then had to explain their choice, based on considerations such as the time it would take to look for a particular concept and how to use a given library. At the beginning of the questionnaire, participants were given two examples to clarify expectations. Responses were collected until $5$ participants provided acceptable answers. To ensure quality, we manually reviewed responses, verifying that participants provided meaningful explanations (e.g., avoiding one-word answers for reasoning questions or random difficulty/time estimations). In particular, while we rely on the time estimation for our analysis, we still ask for the subjective difficulty. Analysing the trend between the two variables for each participant might help in potentially detecting incoherence (e.g., Easy tasks with 40 minutes estimation and Hard tasks with 10 minutes), which could signal random filling of the questionnaire. We followed Alhamed et al. \cite{Alhamed21} procedure until all tasks had been assessed with substantial agreement. We then analysed the difference between human-assessed difficulty and the difficulty for CodeLLMs obtained using our IRT-based approach. Since human assessments are discrete, we discretised the difficulty scores obtained via IRT within the interval [0,1] for comparison.

Applying the above process resulted in a total of \textbf{72} participants. We used prolific filters to only consider developers with at least 3 years of experience. During the questionnaire, the participants were further asked to report their experience using three categories: < 5 years, 5 - 10 years, > 10 years. The experience of the whole pool of the participants is as follows: 33,33\% (24) of the participants have < 5 years of experience, 44,44\% (32) have between 5 and 10 years and 22,22\% (16) have over 10 years.

\section{Results}\label{sec:results}

\subsection{Validation}\label{sec:val}

Before analysing our results, we first proceed to validate the generated prompts. To validate our generated prompts, on top of a manual check, we first inspect the semantic similarity of the prompts rephrasing within and across levels. To do so, we used BERTScore \cite{zhang2019bertscore}, a semantic similarity metric leveraging the cosine similarity of the embedding of references/candidate solutions through a transformer. We did so for each rephrasing against the original prompt of the same level and of other levels. The obtained median similarity for \textit{HumanEval+} ranges from 0.60 (rephrasing of Level 3) to 0.71 (rephrasing of Level 1), and for \textit{ClassEval} ranges from 0.60 (rephrasing of Level 2 and Level 3) to 0.65 (rephrasing of Level 1). For some rephrasing, similarity can drop as low as 0.38 (\textit{HumanEval}+) or 0.28 (\textit{ClassEval}). Inspecting these cases, we did not observe semantic discrepancies. However, the rephrasing obtained via GPT can sometimes be quite convoluted, which might artificially push semantic similarity down, on top of BERTScore's inherent limitation \cite{hanna2021fine}. For instance, \enquote{Clears the expression of all characters that are not brackets.} and \enquote{Retain only bracket characters in the expression, removing all others.} leads to a similarity score of 0.28 despite both sentences expressing the same idea. To further evaluate the validity of the samples, we compared the average similarity within and across levels between rephrasing for the same task using the Wilcoxon test \cite{wilcoxon1992individual}, a nonparametric statistical test evaluating the null hypothesis that two paired samples come from the same distribution. In all cases, we obtained a statistically significant difference (p-value < 0.01) with a strong effect size (> 0.8) \cite{fritz2012effect} when testing whether the distribution of same-level rephrasings was greater than the across-level rephrasings. This highlights that same-level rephrasings are semantically closer to each other than they are to other rephrasings of other levels for the same task.

To further verify the validity of the prompts, we retrained independent IRT models for both datasets using only the aggregated score of code samples of a single level, resulting in 6 different IRT models (3 for each dataset). We would expect that the difficulty of the tasks will drop as the prompt level of information increases. For \textit{HumanEval+}, we found that the median difficulty of the tasks went from 0.42 (Level 1) to 0.15 (Level 3). Similarly, for \textit{ClassEval}, the median difficulty of the tasks dropped from 0.27 (Level 1) to 0.05 (Level 3). Further, using the statistical test of Wilcoxon showed a statistically significant difference (p-value < 0.01) with effect size ranging from 0.25 to 0.64 \cite{fritz2012effect}, when comparing the difficulty for the same task across IRT models of different levels. The lowest effect size was obtained when comparing the difficulty of tasks between Level 1 and Level 2, and the highest was obtained when comparing Level 1 and Level 3. This validates that a higher level of information results in lower difficulty of the tasks, as LLMs get increasingly more information, which allows for more correct samples to be generated.

\subsection{RQ1: \rqone}\label{sec:rqone}

\begin{figure}
\centering
\includegraphics[width=\textwidth]{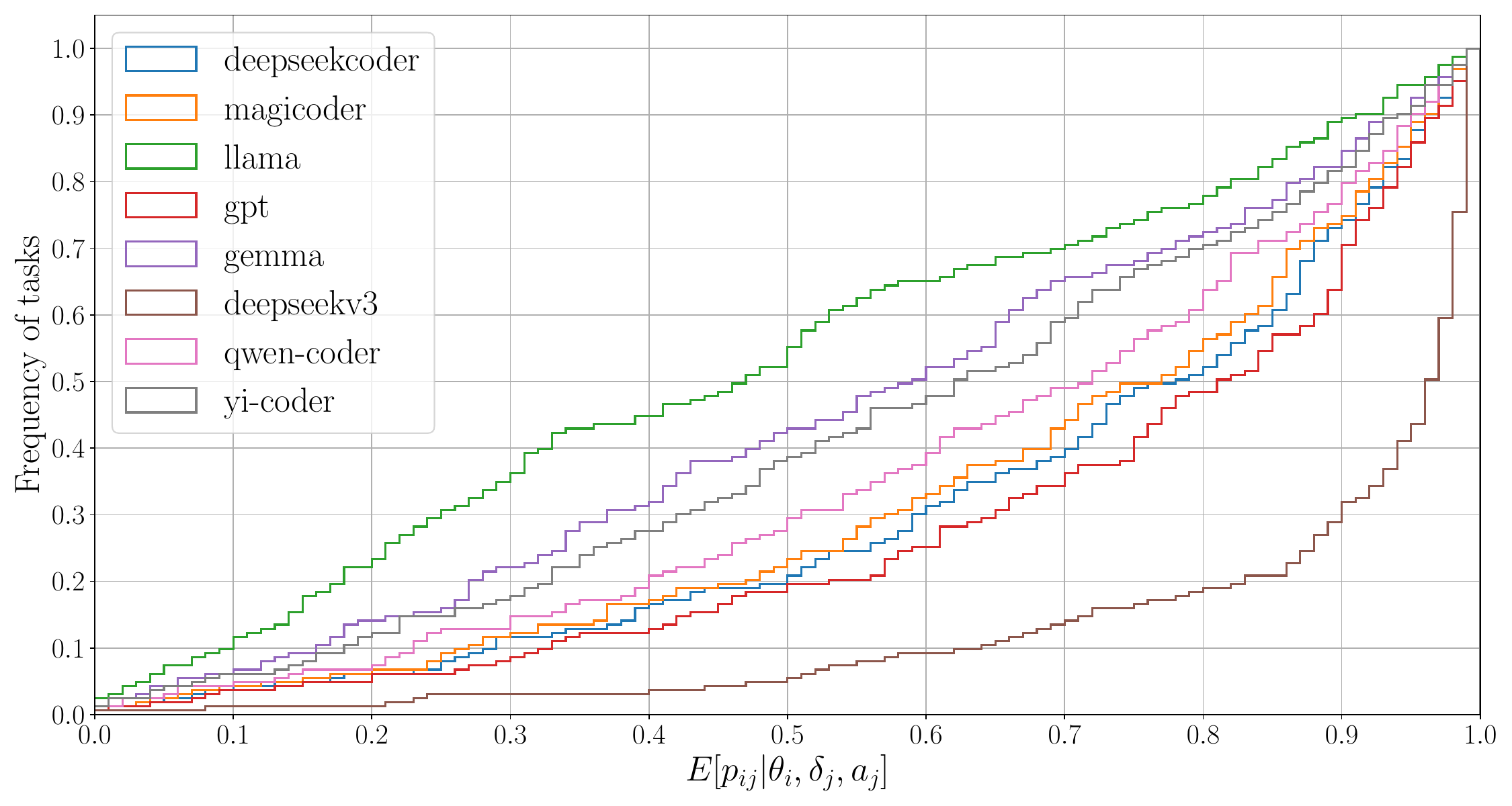}
\includegraphics[width=\textwidth]{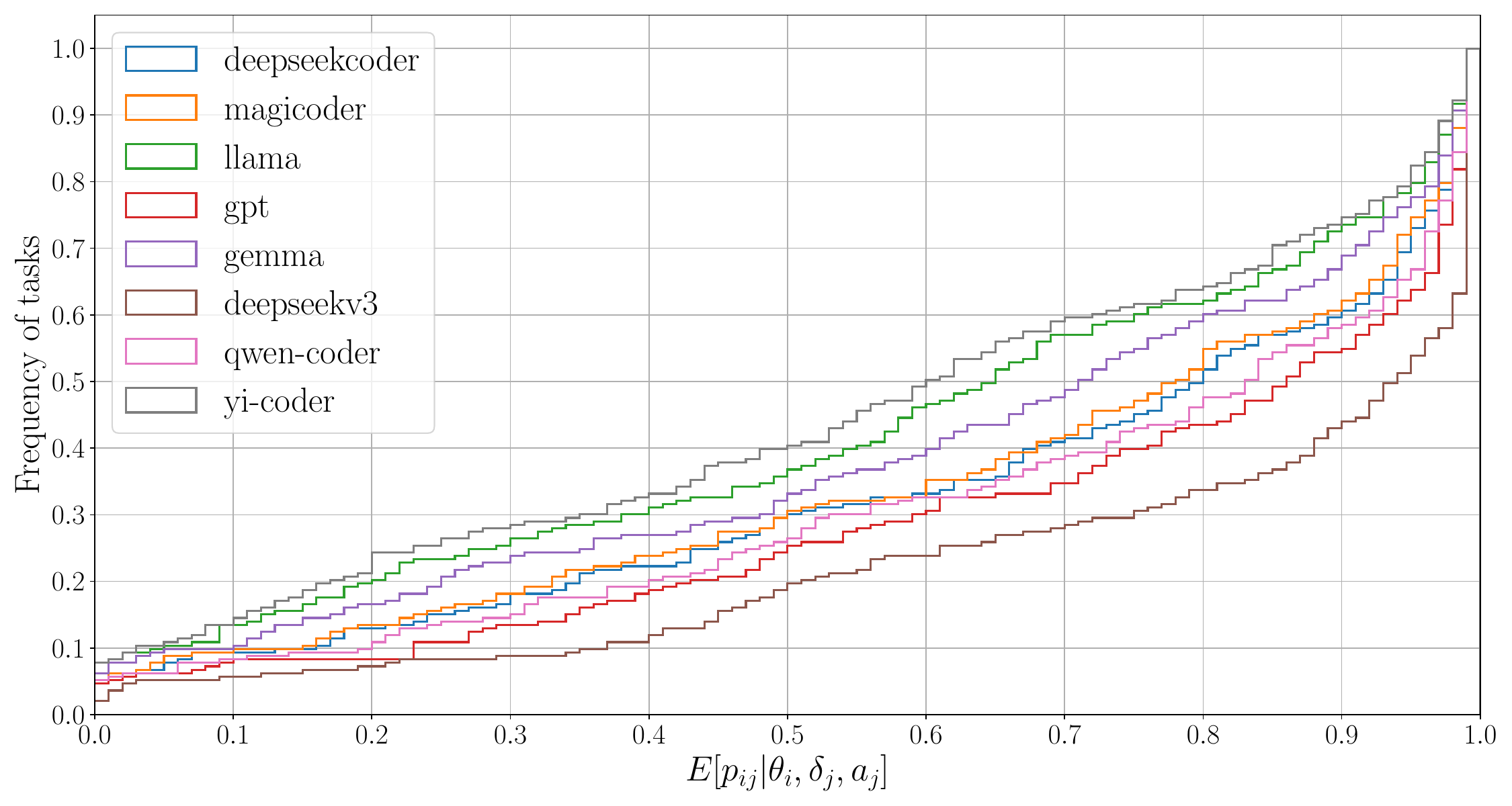}
\caption{Proportion of tasks in \textit{HumanEval+} (\textbf{Top}) and \textit{ClassEval} (\textbf{Bottom}) with a given expected probability of response for each CodeLLM.}
\label{fig:resX-1}
\Description[Cumulative distribution of HumanEval+ and ClassEval frequency of tasks with respect to expected answers differs]{Two figures (top/bottom). In each figure, for each of the eight models used, a cumulative distribution is drawn. More capable models such as DeepSeekV3 tend to have higher frequency of tasks with high expected probability of answering correctly. There are wider variations among models in HumanEval+ than ClassEval. HumanEval+ tends to have more medium difficulty tasks than ClassEval that tends to have more extreme (very easy/difficult) tasks.}
\end{figure}

We first calculated the cumulative distribution of the expected probability of having a correct answer, based on IRT modelling, for each task in each benchmark. The results are presented in Figure \ref{fig:resX-1} and can be read as follows: for instance, for \textit{HumanEval+}, $30\%$ of the tasks for \textit{DeepSeek-V3} have an expected probability of a correct answer of $90\%$ or less (and so $70\%$ have probability $90\%$ or more). These graphs allow us to compare how, globally, tasks are handled by CodeLLMs in both benchmarks. The first observation we can make is regarding the discrepancies between CodeLLMs inside the same benchmark. In the case of \textit{HumanEval+}, there are wider variations between CodeLLMs compared to \textit{ClassEval}. Notably, \textit{Code Gemma}, \textit{Code LLama}, and \textit{Yi-Coder} seem to struggle more on \textit{HumanEval+} with less than $\sim 50\%$ chance of generating a correct code on over $40\%$ of the tasks, compared to the other CodeLLMs. In a second time, we can observe the difference for the same CodeLLM across benchmarks. We observe that \textit{ClassEval} tends to have more extreme tasks compared to \textit{HumanEval+}. Indeed, there are more very easy ($90\%$ or more probability of correct answer) and very hard tasks ($10\%$ or less probability of correct answer) for \textit{ClassEval} compared to \textit{HumanEval+} for most CodeLLMs. On the contrary, \textit{HumanEval+} have more, on average, medium tasks (probability of correct answer between $40\%$ and $60\%$). The difference tends to narrow for models extreme, mainly \textit{CodeLLama} and \textit{DeepSeek-V3}, on \textit{ClassEval} compared to \textit{HumanEval+}. Overall, for the same CodeLLM, there is a difference among benchmarks in terms of the expected probability of answer: computing the AD test returns a statistically significant difference (p$-value < 0.05$) in all cases.

\begin{figure}
\centering
\includegraphics[width=0.49\textwidth]{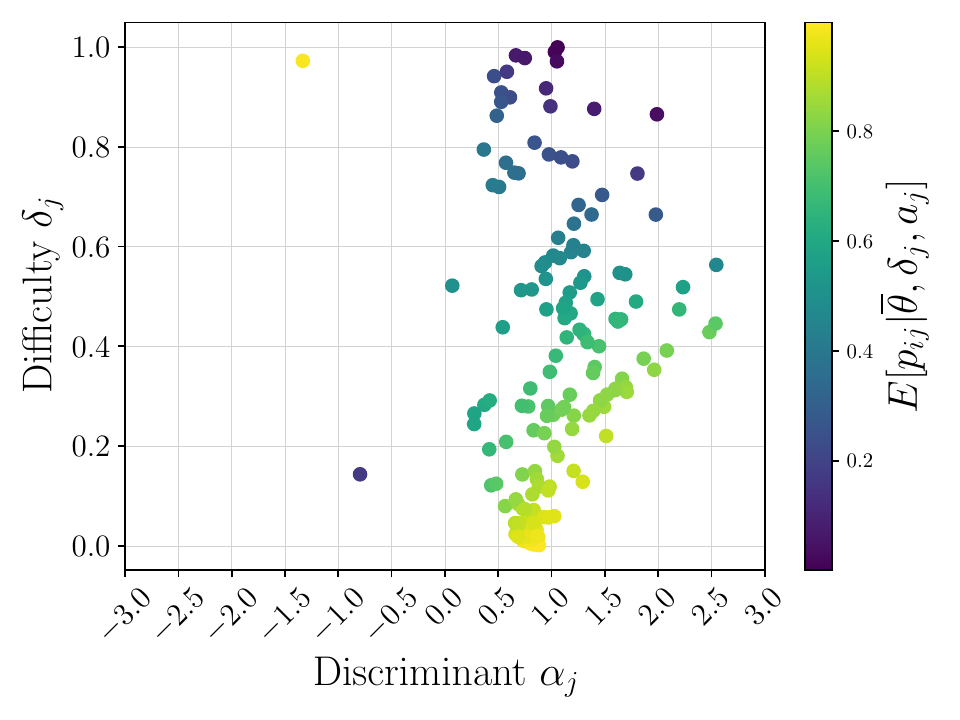}
\includegraphics[width=0.49\textwidth]{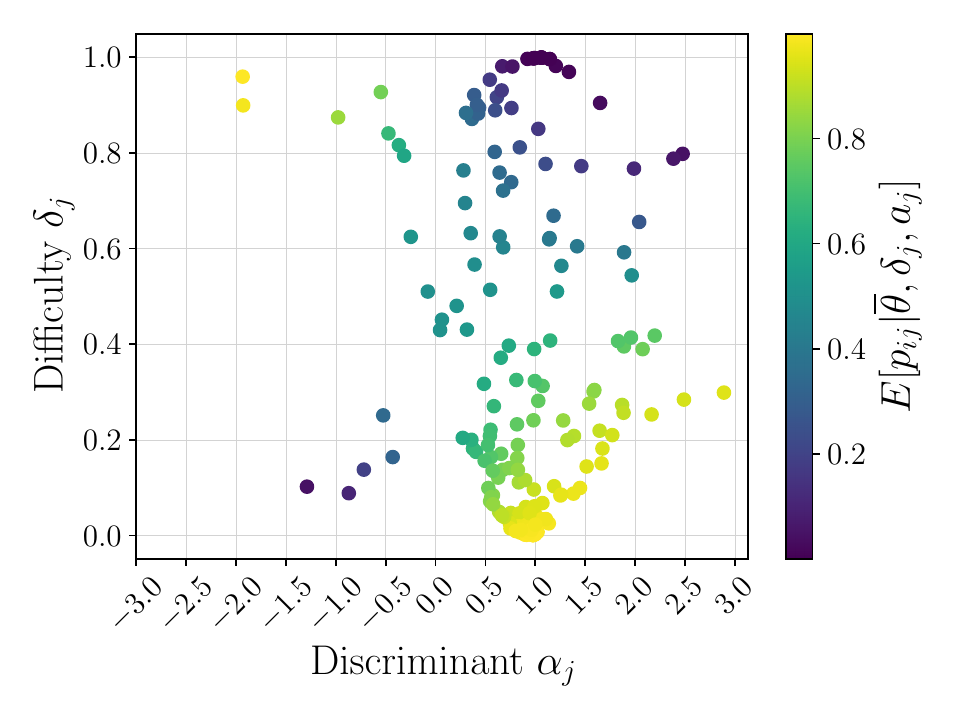}
\caption{Maps of the difficulty $\delta_j$ vs discriminant $a_j$ of each task for a give benchmark. The colour represents the expected probability obtained on a given task by a hypothetical CodeLLM whose capacity $\overline{\theta}$ is the average of the capacity of our CodeLLMs: \textbf{(Left)} HumanEval+ tasks, \textbf{(Right)} ClassEval tasks.}
\label{fig:resX-2}
\Description[HumanEval+ and ClassEval exhibit similar patterns when plotting difficulty against discriminant of tasks]{Two figures (left/right). In each figure, the difficulty against the discriminant of the tasks when considering a CodeLLM with an average capacity are drawn. The higher the difficulty, the less likely the model will answer correctly. Some tasks with negative discriminants are symptomatic of possible annotations errors or specificity of the task.}
\end{figure}

We now present the mapping of difficulty and discriminant parameters for each task in both benchmarks, as computed by the IRT models, in Figure \ref{fig:resX-2}. For each task, we also report the expected probability $E[p_{ij} | \overline{\theta}, \delta_j, a_j]$ as predicted by the model for a hypothetical CodeLLM with a capacity $\overline{\theta}$ as the average of all our CodeLLMs. For both benchmarks, CodeLLMs obtained capacity $\theta_i$ follow the outcome we could expect: \textit{DeepSeek-V3} is the most capable CodeLLM, followed by \textit{GPT3.5}, then \textit{DeepSeekCoder}, \textit{MagiCoder}, and \textit{QwenCoder} are tied together, and finally \textit{Code Gemma}, \textit{Code Llama}, and \textit{Yi-Coder} are similar. From the mapping, on the positive discriminant side, we see that the difficulty correlates with the expected probability: intuitively, the higher the difficulty of the task, the smaller the probability of the answer should be. We also observe some \enquote{peak} of high discriminability around tasks of difficulty between $0.4$ and $0.6$ for \textit{HumanEval+} and around $0.2$ and $0.4$ as well as $0.6$ and $0.8$ for \textit{ClassEval}: the most discriminative tasks, allowing to tell apart which CodeLLMs are better, are more likely to not to be in the extreme of difficulty. Indeed, for very hard tasks or very easy tasks, either all CodeLLMs fail or all CodeLLMs manage in a similar proportion, which makes it difficult to discriminate between them. This echoes what we observed with the cumulative distribution of the expected probability of answer in Figure \ref{fig:resX-1}, with a wider discrepancy between CodeLLMs.

We note, however, an inverse pattern in terms of difficulty / expected probability when we reach the tasks with a negative discriminant. It should be noted, from Figure \ref{fig:ex_irt}, that a negative discriminant means that less capable CodeLLMs are more likely to have a correct answer than more capable CodeLLMs. We make the distinctions between two parts of this side of the diagram: the tasks with high difficulty and low difficulty, as symbolised by the $0.5$ threshold. For tasks with negative discriminant and high difficulty, we also note that CodeLLMs tend to have a relatively high expected probability on these tasks. In that case, this can be explained by both non-determinism in the prompts and the small number of CodeLLMs used. For instance, in the green curve in Figure \ref{fig:ex_irt}, if all CodeLLMs have abilities between $0.3$ and $0.5$, the difference between expected probabilities would be small, but the IRT models would yield a negative discriminant with a high difficulty. We noted this behaviour when realising our experiments: computing the difficulty/discriminant with a subset of models compared to the whole set resulted in some increase of the number of tasks with a negative discriminant, with the magnitude of the increase varying depending on the subset chosen. For instance, when using models \{\textit{deepseekcoder}, \textit{magicoder}, \textit{codellama}, \textit{codegemma}, \textit{gpt}\} only, we observed an increase from 1.2\% to 8.5\% on HumanEval+ and 9.5\% to 15\% on ClassEval of the proportions of such tasks compared to using all models.

The second part of this half of the diagrams has tasks with low difficulty and a negative discriminant. In that case, most CodeLLMs will fail such tasks. This is generally symptomatic of possible annotation errors, as less capable CodeLLMs do better than more capable ones, but not all CodeLLMs are generally good. We thus manually investigated these tasks. Among the $9$ tasks from \textit{ClassEval} and $1$ task from \textit{HumanEval+}, we found issues in the way the original prompts (and so our prompts generated from them) were formulated for $7$ of them in \textit{ClassEval} and $1$ in \textit{HumanEval+}. For instance, in ClassEval, one task consists of implementing the cosine function using a Taylor approximation. We give in Listing \ref{list:example-bad-1} an excerpt with relevant information. From the excerpt, the CodeLLMs have access to the signature of the function implementing the `taylor` function, but not the actual implementation. The oracle implementation contains code to convert $x$ into radians before returning the approximation. This poses an issue as, without this information, most CodeLLMs will generate codes in the `cos` that convert $x$ into radians before using the `taylor` function, leading to an error. \textit{Code LLama} and \textit{Code Gemma} tend not to follow this approach as much, hence why the a negative discriminant and a low expected probability.

\begin{lstlisting}[language=python, caption={Excerpt from the `cos` task of \textit{ClassEval} following the original prompt}, label={list:example-bad-1}]
from math import pi, fabs
class TriCalculator:  
    """
    The class allows to calculate trigonometric values, including cosine, sine, and tangent, using Taylor series approximations.
    """
    def __init__(self):
        pass
    ...
    def taylor(self, x, n):
        pass # Implementation not given, but oracle function converts x into radian
    ...
    def cos(self, x):
        """
        Calculate the cos value of the x-degree angle
        :param x:float
        :return:float
        >>> tricalculator = TriCalculator()
        >>> tricalculator.cos(60)
        0.5
        """

\end{lstlisting}

On the other hand, for the tasks we investigated where no issues were detected, we observed behaviours that may reveal specific characteristics of CodeLLMs, possibly influenced by their training data. Given in Listing \ref{list:example-bad-2} is the original prompt for a task of \textit{HumanEval+} where the goal is to filter a list to retain only integers. While relatively simple, more able CodeLLMs like \textit{GPT3.5} fail several times on it when \textit{Code LLama} is better at managing it. Looking at the answers provided, the codes generated are similar, except that \textit{GPT3.5} generally relies on `isinstance` while \textit{Code LLama} relies on `type` to check for integers. While `isinstance` is generally more useful as it accounts for inheritance, however, in that case, `type` should be used. Indeed, if the list were, for instance, to contain boolean `True` or `False`, `isinstance` would consider them as integers (as the boolean type inherits from integer in Python) while `type` will not. This can be indicative of the specificity of the training dataset or, in the worst case, some form of memorisation.
 
\begin{lstlisting}[language=python, caption={Excerpt from the `filter\_integers` task of \textit{HumanEval+} following the original prompt}, label={list:example-bad-2}]
from typing import List, Any

def filter_integers(values: List[Any]) -> List[int]:
    """
    Filter given list of any python values only for integers
    >>> filter_integers(['a', 3.14, 5])
    [5]
    >>> filter_integers([1, 2, 3, 'abc', {}, []])
    [1, 2, 3]
    """    

\end{lstlisting}

\begin{tcolorbox}[colback=blue!5,colframe=blue!40!black]
\textbf{Findings 1:} IRT allows us to map tasks in terms of difficulty and discriminant and to contrast benchmarks. CodeLLMs have varying behaviour within and between benchmarks: \textit{ClassEval} tends to have more extreme (very easy/difficult) tasks while \textit{HumanEval+} have more medium tasks in terms of difficulty for the CodeLLMs under study, with more discrepancies in terms of expected answers in \textit{HumanEval+}. Most discriminant tasks in both benchmarks generally have a difficulty between $0.2$ and $0.8$, which thus are better at separating CodeLLMs. Few tasks have negative discriminant, symptomatic of small differences due to non-determinism or annotation errors.
\end{tcolorbox}

\subsection{RQ2: \rqtwo}\label{sec:rq2}

\begin{figure}[h!]
\centering
\includegraphics[width=0.95\textwidth]{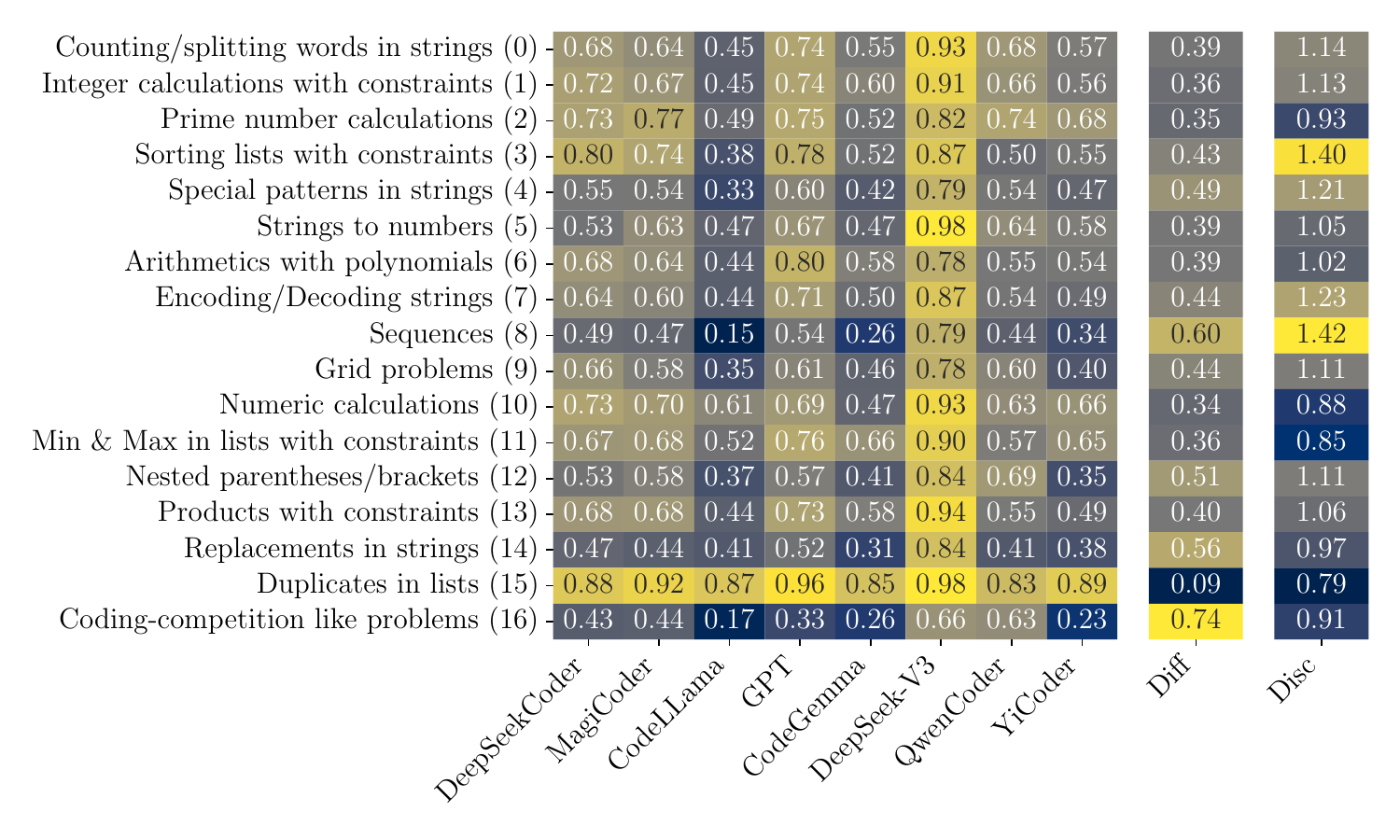}
\break   
\includegraphics[width=0.95\textwidth]{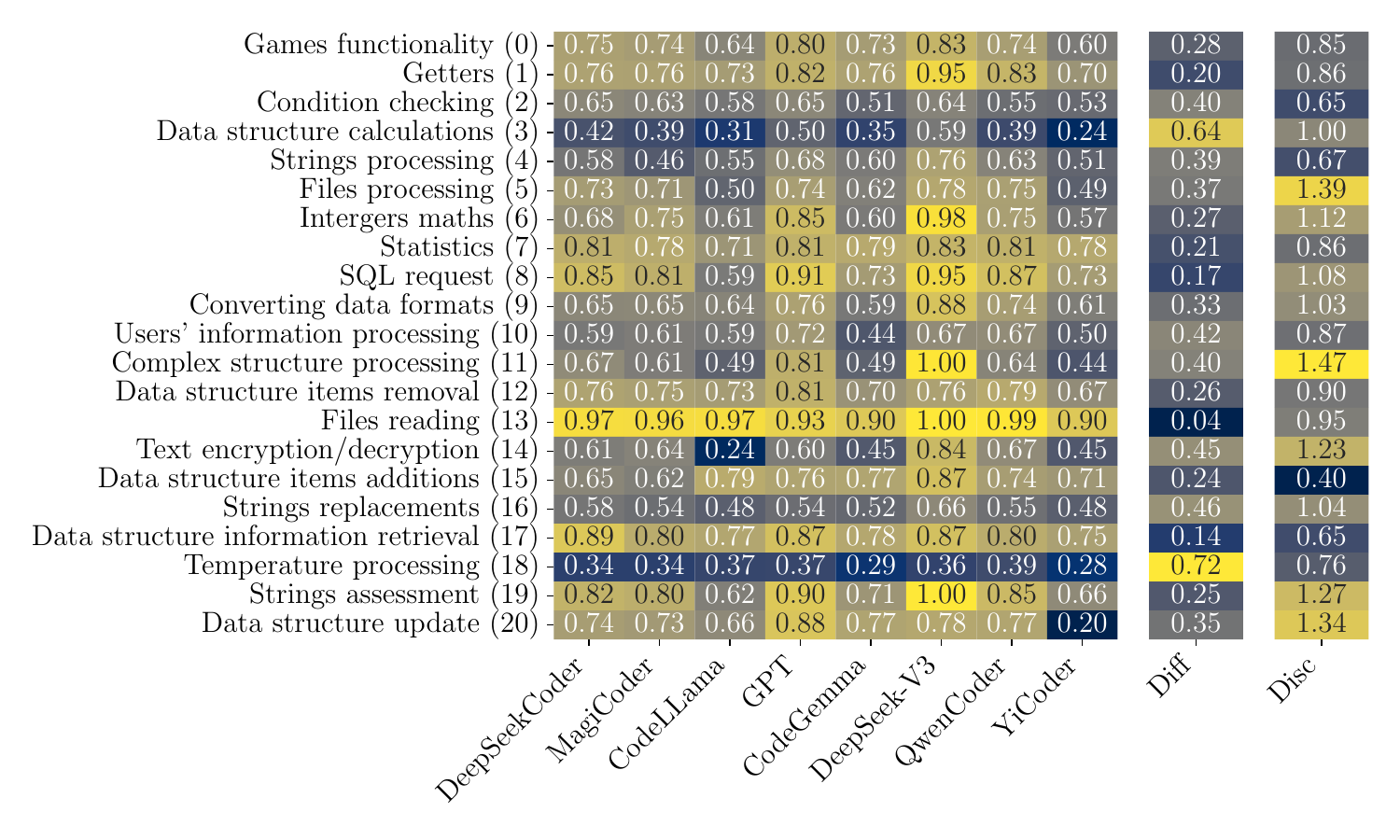}
\caption{Mean per topic of: 1) the accuracy on the tasks' prompts for each CodeLLM, 2) the tasks' difficulty (\textit{Diff}), 3) the tasks' discriminant (\textit{Disc}). (\textbf{Top}) \textit{HumanEval+}, (\textbf{Bottom}) \textit{ClassEval}.}
\label{fig:topics}
\Description[Difficulty/Discriminant of extracted topics varies for both ClassEval and HumanEval+]{Two figures (top/bottom). In each figure, for each the topic extracted, the expected answer per model as well as the average difficulty/discriminant per task is calculated. For each benchmark, we note some difficult/discriminant topics that could be use to further probe the models, as they highlight shortocmings of models.}
\end{figure}

In this section, we proceed with the exploration of the topics. Results are given in Figure \ref{fig:topics}. Starting with \textit{HumanEval+}, we see that most topics have a mean difficulty in the easy-medium range ($0.3 - 0.5$). The last two topics show a high variation from the rest of the topics; however, this might be explained by the lower number of individual tasks in these topics, making them particularly specialised. Out of the other topics, four of them have a mean difficulty above $0.45$: \textit{Nested parentheses/brackets} (12), \textit{Sequences} (8), \textit{Special patterns in strings} (4) and \textit{Replacements in strings} (14). \textit{Sequences} (e.g., Fibonacci) also have a high positive discriminant, pointing out that less capable CodeLLMs such as \textit{Code Llama} tend to have a hard time processing those tasks. Such tasks can be interesting in distinguishing between different CodeLLMs. The mean accuracy of the CodeLLMs varies from topic to topic, but \textit{DeepSeek-V3} is generally the model with the highest mean accuracy and \textit{Code Llama} the one with the lowest. Most topics have a mean discriminant around $1$, and no topics have a negative mean discriminant. Such topics are more likely to distinguish between more/less capable CodeLLMs, though there can be some variability between similarly capable CodeLLMs such as \textit{MagiCoder}, \textit{Deepseek} and \textit{QwenCoder} or \textit{Code Llama}, \textit{Code Gemma} and \textit{YiCoder}.

Regarding \textit{ClassEval}, we see that, contrary to \textit{HumanEval+}, most topics are fairly easy, with $48\%$ of the topics with a mean difficulty below $0.3$ and most CodeLLMs achieving a higher mean accuracy compared to \textit{HumanEval+}. As such, \textit{SQL Requests} (8), \textit{Files reading} (13) (e.g., opening a JSON file and returning the content), and \textit{Data Structure information retrieval} (17) (e.g., finding data in a dictionary) are relatively trivial for all CodeLLMs. On the contrary, besides \textit{Temperature processing} (18), which might have a very high difficulty due to a low number of very specialized tasks (dealing with classes with functions related to ambient temperature control), three topics have a mean difficulty above $0.45$: \textit{Data structure calculations} (3) (e.g., computing Pearson correlations between two lists),  \textit{Text encryption/decryption} (14) (e.g., implementing Caesar cipher), and \textit{Strings replacements} (16) (e.g., replacing a character by another in a string). Compared to \textit{HumanEval+}, there are more low discriminant topics (discriminant $< 1$ or even $< 0.75$), which highlights that \textit{ClassEval} contains more tasks where CodeLLMs tend to obtain similar results. As we mentioned in Section \ref{sec:rqone}, there were fewer discrepancies between CodeLLMs in terms of the expected probability of answer in \textit{ClassEval} than \textit{HumanEval+}. Within these topics, we can even see some of them for which less capable CodeLLMs (\textit{Code Gemma} and \textit{Code Llama}) achieve a higher mean accuracy than more capable ones: it is the case for \textit{String processing} (4) (e.g., removing stop-words from a text with NLTK) or \textit{Data structure items additions} (15). In the case of \textit{Data structure items additions} (15) \textit{Code Llama} and \textit{Code Gemma} are, on average, better than \textit{QwenCoder} or \textit{GPT3.5}. Analysing the tasks within the topic, we observe that more capable models often infer additional assumptions about the task, leading to unnecessarily complex generated code. As an example, we give the generated codes for one such task of \textit{GPT3.5} and \textit{Code Llama} along with the original prompt in Listing \ref{list:example-3}.

\begin{lstlisting}[language=python, caption={Example of discrepancy on a task from Topic 15. We give first one of our prompt for the task, then generated code by GPT3.5, and CodeLlama.}, label={list:example-3}]
# Original prompt
from datetime import datetime
class Classroom:  
    """
    This is a class representing a classroom, capable of adding and removing courses, checking availability at a given time, and detecting conflicts when scheduling new courses.
    """

    def __init__(self, id):
        """
        Initialize the classroom management system.
        :param id: int, the id of classroom
        """
        self.id = id
        self.courses = []

    def check_course_conflict(self, course):
        pass
        
    ... # Some other class functions

    def add_course(self, course):
        """Add a course to the "self.courses" list if the course is not already in it.
        :param course: dict, information of the course, including 'start_time', 'end_time' and 'name'
        """
# GPT3.5 code
def add_course(self, course):
    if course not in self.courses and not self.check_course_conflict(course):
        self.courses.append(course)

# Code Llama code
def add_course(self, course):
    if course not in self.courses:
        self.courses.append(course)
\end{lstlisting}

As one can see, \textit{GPT3.5} assumes there is a need to check for some conflicts, possibly due to the `course` parameter containing information such as `start\_time` and the function `self.check\_course\_conflict` being available as a class function from the prompt. However, this is not requested at all in the prompt (and in the task), which is quite straightforward. On the contrary, \textit{Code Llama} manages the task by following exactly the instructions mentioned in the prompt. We observe a similar phenomenon for other generated code on the same task. 

\begin{tcolorbox}[colback=blue!5,colframe=blue!40!black]
\textbf{Findings 2:} Analysing topics can reveal variations in how CodeLLMs handle different types of tasks. Some topics are notably harder for CodeLLMs in both benchmarks (e.g., \textit{Sequences} and \textit{Nested parentheses/brackets} in \textit{HumanEval+} and \textit{Data structure calculations} in \textit{ClassEval}), while some are trivial (e.g., \textit{File Reading} in \textit{ClassEval}). Discriminant information gives us clues on some topics of interest, highlighting the tasks' topics where CodeLLMs can be more or less easily separated depending on their capabilities.
\end{tcolorbox}

\subsection{RQ3: \rqthree}

Results of the correlations between node types extracted and tasks' characteristics, as well as AD test are given in Table \ref{tab:struct-he} and Table \ref{tab:struct-he-disc} for \textit{HumanEval+}, and in Table \ref{tab:struct-ce} for \textit{ClassEval} (we did not find any significant correlation between discriminants and program constructs for \textit{ClassEval}). To facilitate interpretation, we regrouped node types into semantically similar categories before calculating the categories. Hence, \textit{Op} includes all nodes related to math operations such as addition, subtraction or even boolean operations, \textit{Loop} includes both \textit{For} and \textit{While} node types, \textit{DataStruct} contains all \textit{List}, \textit{Set}, \textit{Dict} or \textit{Tuple} node types, etc. All details on the grouping process are available in our replication package~\cite{rep-package}.

Observing the results, we distinguish several patterns that can be intra-dataset (across CodeLLMs) or inter-dataset (between \textit{HumanEval+} and \textit{ClassEval}). First, looking at the comparison with tasks' difficulty (Table \ref{tab:struct-he} and Table \ref{tab:struct-ce}), several node types such as \textit{If} and \textit{Assign} are present across CodeLLMs and benchmarks, highlighting a commonality in the programming constructs that CodeLLMs use to address harder/easier tasks. While harder tasks could result in longer code, which might bias the outcome, as we controlled for the size of the code fragments, it points out that it is not simply about more absolute processing, but that to tackle harder task,s CodeLLMs rely on a higher level of processing. Moreover, since the trend is observed at different similarity thresholds, this points towards an intrinsic property of easy/hard tasks. As such, the harder the tasks, the more the CodeLLMs need conditions (especially comparison operators, given the widespread significance of \textit{Compare} in both benchmarks) independently of the number of lines of code to solve the task. Additionally, increasingly harder tasks require more manipulation of data through variable assignment. In turn, this might increase their chance of producing erroneous code as they get lost in the logic flow. Note that this highlights how CodeLLMs attempt to solve the task, rather than what is \textit{required} to solve the task. There are various ways to approach each task, and optimising the code was not a requirement for CodeLLMs. 

In contrast, some differences emerge between the two benchmarks, highlighted by stronger correlations and/or more significant results from the AD test for CodeLLMs. For \textit{HumanEval+}, CodeLLMs tend to utilise more \textit{DataStruct} (e.g., list, dict etc.) constructs when tackling harder tasks. For \textit{ClassEval}, CodeLLMs make more use of index manipulation (\textit{Subscript}) or operators (\textit{Op} such as boolean 'or', binary 'not' or simple addition '+'). In that last case, we find non-significant correlations for this same node type for \textit{HumanEval+}, despite said node type being used in \textit{HumanEval+} tasks. In that case, it means that this is not a property of tasks in general, but it is a property of the tasks of \textit{ClassEval}. Correlations are mostly \textit{weak} ($> 0.06$), with only the correlations with the \textit{If} node in \textit{HumanEval+}/\textit{ClassEval} and \textit{Op} for \textit{ClassEval} being \textit{moderate} ($> 0.26$) \cite{gilpin1993table, schober2018correlation}. It should be noted that it is possible to have significant AD test without significant correlations (e.g., \textit{Lambda} in \textit{HumanEval+}) or the inverse (e.g., \textit{arg} in \textit{ClassEval}). This can arise when there are differences between the distribution of the top/bottom 50\% tasks' difficulty, but without a clear trend in terms of difficulty, and so no correlations. 

Regarding CodeLLMs, we observe that for some node types, some CodeLLMs do or do not exhibit a correlation/AD difference, while the other CodeLLMs exhibit the opposite. For instance, in \textit{HumanEval+}, \textit{FunctionDef} nodes exhibit a correlation for all but \textit{DeepSeek-V3} and \textit{QwenCoder}, highlighting that these models might not rely on ancillary functions to solve HumanEval's harder tasks. In \textit{ClassEval}, \textit{DeepSeek-V3} is the only CodeLLM without a significant correlation/AD for \textit{comprehension} (e.g. list comprehension, etc.). This highlights that CodeLLMs will deal with tasks within the benchmark differently, based on their understanding of the prompt, their training dataset, and their architecture. While this can lead to errors, as we saw for instance in Listing \ref{list:example-3} in the previous Section \ref{sec:rq2}, it can also be harmless but emphasise particularity in how the prompts are processed.

\begin{table}[h]
    
    \centering
    \caption{Top program construct (node type) for each CodeLLM with regard to the tasks' difficulty for HumanEval+. $\tau$ gives the Kendall$-\tau$ correlations (lowest/highest among thresholds) between the program constructs used in each CodeLLM and the tasks' difficulty. $AD$ gives the result of the Anderson-Darling test between the programming constructs, the top $50\%$ easiest tasks vs the top $50\%$ hardest tasks. Significance is calculated at $p-value < 0.05$ using the Holm-Bonferroni correction.}
    \label{tab:struct-he}
    \resizebox{\textwidth}{!}{\begin{tabular}{c*{7}{cc|}cc}
    \toprule   
         \multirow{3}{*}{Node} & \multicolumn{16}{c}{CodeLLMs} \\
         \cline{2-17}
         & \multicolumn{2}{c|}{DeepSeekCoder} & \multicolumn{2}{c}{MagiCoder} & \multicolumn{2}{c|}{CodeLlama} & \multicolumn{2}{c|}{GPT3.5} & \multicolumn{2}{c|}{CodeGemma}& \multicolumn{2}{c|}{DeepSeek-V3}& \multicolumn{2}{c|}{QwenCoder} & \multicolumn{2}{c}{YiCoder} \\ \cline{2-17}
         & $\tau$& AD & $\tau$& AD & $\tau$& AD & $\tau$& AD& $\tau$& AD& $\tau$& AD& $\tau$& AD& $\tau$& AD \\
         \midrule
         \textit{Assign} & .24/.26 & * & .22/.25 & * & .25/.27 & * & .22/.24 & * & .18/.22 & * & .26/.28 & * & .22/.24 & * & .23/.25 & *  \\ 
        \textit{Lambda} & - & * & - & * & - & - & - & * & - & - & - & - & - & - & - & -  \\ 
        \textit{Subscript} & .14/.16 & * & .15/.17 & * & .14/.16 & * & - & - & .14/.16 & - & - & - & .13/.15 & - & .17/.18 & *  \\ 
        \textit{If} & .27/.30 & * & .26/.29 & * & .24/.27 & * & .27/.31 & * & .25/.29 & * & .28/.30 & * & .26/.30 & * & .26/.31 & *  \\ 
        \textit{Loop} & .12/.14 & - & .13/.16 & * & .13/.16 & - & .13/.16 & * & - & - & .13/.15 & - & .14/.15 & * & .12/.16 & -  \\ 
        \textit{Compare} & .14/.16 & * & .12/.15 & * & - & - & .13/.16 & * & .11/.14 & * & .10/.12 & * & .13/.14 & * & .14/.15 & *  \\ 
        \textit{comprehension} & - & * & - & - & - & - & - & - & - & * & - & - & - & - & - & -  \\ 
        \textit{FunctionDef} & .16/.25 & * & .13/.21 & * & .13/.27 & * & .15/.20 & * & .13/.23 & * & - & - & - & - & .13/.24 & *  \\ 
        \textit{DataStruct} & .15/.19 & * & .14/.18 & * & .17/.21 & * & .14/.17 & * & .15/.19 & * & .17/.18 & * & .13/.14 & * & .13/.17 & *  \\         
    \bottomrule
    \end{tabular}}
\end{table}

\begin{table}[h]
    
    \centering
    \caption{Top program construct (node type) for each CodeLLM with regard to the tasks' difficulty for ClassEval. $\tau$ gives the Kendall$-\tau$ correlations (lowest/highest among thresholds) between the program constructs used in each CodeLLM and the tasks' difficulty. $AD$ gives the result of the Anderson-Darling test between the programming constructs, the top $50\%$ easiest tasks vsthe  top $50\%$ hardest tasks. Significance is calculated at $p-value < 0.05$ using the Holm-Bonferroni correction.}
    \label{tab:struct-ce}
    \resizebox{\textwidth}{!}{\begin{tabular}{c*{7}{cc|}cc}
    \toprule   
         \multirow{3}{*}{Node} & \multicolumn{16}{c}{CodeLLMs} \\
         \cline{2-17}
         & \multicolumn{2}{c|}{DeepSeekCoder} & \multicolumn{2}{c}{MagiCoder} & \multicolumn{2}{c|}{CodeLlama} & \multicolumn{2}{c|}{GPT3.5} & \multicolumn{2}{c|}{CodeGemma}& \multicolumn{2}{c|}{DeepSeek-V3}& \multicolumn{2}{c|}{QwenCoder} & \multicolumn{2}{c}{YiCoder} \\ \cline{2-17}
         & $\tau$& AD & $\tau$& AD & $\tau$& AD & $\tau$& AD& $\tau$& AD& $\tau$& AD& $\tau$& AD& $\tau$& AD \\
         \midrule
         \textit{arg} & .14/.16 & * & .11/.14 & - & - & - & .12/.15 & - & .11/.15 & - & .12/.14 & - & .13/.16 & - & .12/.13 & -  \\ 
        \textit{Assign} & .14/.24 & - & .17/.27 & * & .17/.26 & * & .19/.29 & * & .12/.24 & - & .16/.25 & * & .14/.25 & * & .18/.29 & *  \\ 
        \textit{Subscript} & .20/.31 & * & .19/.28 & * & .19/.29 & * & .18/.27 & * & .21/.29 & * & .17/.25 & * & .20/.27 & * & .19/.31 & *  \\ 
        \textit{If} & .25/.28 & * & .25/.26 & * & .24/.27 & * & .25/.27 & * & .25/.27 & * & .28/.29 & * & .27/.29 & * & .28/.31 & *  \\ 
        \textit{Loop} & - & - & .16/.24 & * & .12/.25 & - & .14/.23 & - & - & - & .14/.23 & * & .14/.24 & * & .13/.28 & -  \\ 
        \textit{Attribute} & - & - & - & - & - & - & - & - & - & - & - & * & - & - & - & -  \\ 
        \textit{Compare} & .17/.22 & * & .17/.21 & * & .16/.18 & * & .16/.20 & * & .19/.22 & * & .19/.21 & * & .16/.19 & * & .20/.24 & *  \\ 
        \textit{comprehension} & .17/.27 & * & .14/.26 & - & .15/.24 & - & .16/.24 & - & .15/.22 & * & - & - & .14/.24 & * & .12/.24 & -  \\ 
        \textit{DataStruct} & .12/.23 & * & .12/.22 & * & .11/.20 & * & - & - & - & - & - & - & .14/.24 & * & .16/.27 & *  \\ 
        \textit{Op} & .25/.30 & * & .23/.29 & * & .29/.31 & * & .25/.30 & * & .28/.31 & * & .24/.28 & * & .28/.31 & * & .28/.33 & *  \\         
    \bottomrule
    \end{tabular}}
\end{table}

Regarding the comparison with the tasks' discriminant, we note that there are a few cross-CodeLLMs' patterns for \textit{HumanEval+}, namely \textit{Assign}, \textit{Attributes} and \textit{DataStruct}. Both \textit{Assign} and \textit{DataStruct} were already significant for tasks' difficulty. We might explain this behaviour by looking at Figure \ref{fig:resX-2}: as the cut-off for top/bottom 50\% most/least discriminant tasks is around 1, we see that highly discriminant tasks tend to be, on average, more difficult than low-discriminant tasks for \textit{HumanEval+}. As such, we might observe a similar pattern in the case of \textit{HumanEval+} because of this phenomenon. In the case of \textit{HumanEval+}, the more discriminant the task, the more likely variable assignment and data structure will be used. At the same time, the more code samples will contain attribute usage (e.g. \enquote{mylist.append()} where \textit{append} is an attribute of the list). On the other hand, we do not observe any patterns for \textit{ClassEval}.

\begin{table}[h]
    
    \centering
    \caption{Top program construct (node type) for each CodeLLM with regard to the tasks' discriminant for HumanEval+. $\tau$ gives the Kendall$-\tau$ correlations (lowest/highest among thresholds) between the program constructs used in each CodeLLM and the tasks' discriminant. $AD$ gives the result of the Anderson-Darling test between the programming constructs, the top $50\%$ least discriminant tasks vs the top $50\%$ most discriminant tasks. Significance is calculated at $p-value < 0.05$ using the Holm-Bonferroni correction.}
    \label{tab:struct-he-disc}
    \resizebox{\textwidth}{!}{\begin{tabular}{c*{7}{cc|}cc}
    \toprule   
         \multirow{3}{*}{Node} & \multicolumn{16}{c}{CodeLLMs} \\
         \cline{2-17}
         & \multicolumn{2}{c|}{DeepSeekCoder} & \multicolumn{2}{c}{MagiCoder} & \multicolumn{2}{c|}{CodeLlama} & \multicolumn{2}{c|}{GPT3.5} & \multicolumn{2}{c|}{CodeGemma}& \multicolumn{2}{c|}{DeepSeek-V3}& \multicolumn{2}{c|}{QwenCoder} & \multicolumn{2}{c}{YiCoder} \\ \cline{2-17}
         & $\tau$& AD & $\tau$& AD & $\tau$& AD & $\tau$& AD& $\tau$& AD& $\tau$& AD& $\tau$& AD& $\tau$& AD \\
         \midrule
         \textit{Assign} & .15/.19 & * & .15/.18 & * & .15/.20 & * & .13/.17 & * & .14/.19 & * & .13/.18 & * & .17/.20 & * & .15/.19 & *  \\ 
        \textit{Lambda} & - & - & - & * & - & - & - & * & - & - & - & - & - & - & - & -  \\ 
        \textit{Subscript} & - & * & - & * & - & * & - & * & - & * & - & * & - & * & - & *  \\
        \textit{Attribute} & .15/.18 & * & .15/.18 & * & .18/.20 & * & .11/.13 & * & .14/.15 & * & .14/.17 & * & .17/.19 & * & .15/.16 & *  \\ 
        \textit{Compare} & - & * & - & * & - & - & - & - & - & - & - & * & - & - & - & -  \\ 
        \textit{comprehension} & - & - & - & - & - & - & - & - & - & * & - & - & - & - & - & -  \\ 
        \textit{Call} & - & - & - & - & - & - & - & - & - & - & - & * & - & - & - & -  \\ 
        \textit{FunctionDef} & - & - & .14/.19 & * & - & * & .18/.24 & * & - & - & - & - & .13/.18 & * & - & -  \\ 
        \textit{DataStruct} & .11/.15 & * & .13/.15 & * & .14/.18 & * & .13/.15 & * & .16/.18 & * & .15/.17 & * & .11/.13 & * & .14/.16 & *  \\ 
        \textit{Op} & - & * & - & * & - & - & - & * & - & - & - & * & - & * & - & -  \\        
    \bottomrule
    \end{tabular}}
\end{table}

\begin{tcolorbox}[colback=blue!5,colframe=blue!40!black]
\textbf{Findings 3:} Cross-checking the AST of generated codes with tasks' characteristics points towards specific program constructs used by CodeLLMs to address a task. Conditions and variable assignment are generally used more in harder tasks, independently of the code length, while some node types are more specific to a particular benchmark (e.g., \textit{Subscript} in \textit{ClassEval} and \textit{DataStruct} in \textit{HumanEval+}). We do not find any particular patterns when using the discriminant of the tasks for both benchmarks, 
though we observe correlations for the same specific node types for \textit{HumanEval+} (i.e. \textit{Assign}, \textit{Attribute}, etc.).
\end{tcolorbox}

\subsection{RQ4: \rqfour}\label{sec:human_exp}

As detailed in Section \ref{sec:rqs}, estimating the difficulty of tasks has been conducted in several iterative rounds. After the first round, only $6$ tasks did not reach the aimed agreement. In the second round, we thus redid one questionnaire with these $6$ tasks, following the same template, except we added the minimum/median/maximum time/difficulty estimations of the previous round, as well as comments provided by previous participants. We obtained again $5$ answers and computed the agreement, resulting in only $1$ task not reaching the threshold. Analysis of this task revealed that all but one participant judged the task long (40 minutes or more), and one participant judged it quick (5 minutes), which resulted in an agreement below $0.6$ because of the metric used in Alhamed et al. \cite{Alhamed21}, the same as the first round. Given this observation, even with a lower agreement, we accepted the task as most participants agreed on the time effort.

To analyse the results, we first compare the human annotators' difficulty assessment between \textit{HumanEval+} and \textit{ClassEval}. The AD test between both distributions reveals a p$-value < 0.01$, showing a significant difference between the two distributions. When examining the distributions, we found that annotators tended to rate tasks from \textit{ClassEval} more difficult than \textit{HumanEval+}: $40\%$ of \textit{ClassEval} were labelled as taking $20$ minutes or more, while only $17\%$ of \textit{HumanEval+} were labelled as requiring the same time. This contrasts with the assessment made by CodeLLMs, as we saw in RQ1 and RQ,2 that \textit{HumanEval+} tended to have a higher number of medium-difficult tasks for the CodeLLMs compared to \textit{ClassEval}. While this could be attributed to a bias in our sampled tasks, we found no significant difference between the difficulty distributions of our sampled tasks and those of the entire set of tasks.

We then compared the human annotators' difficulty assessment and the CodeLLMs' assessment. Given that we have two types of rating on the same items, we can use the Cohen-$\kappa$ to measure the agreement between the two raters (i.e., the human annotators' median and the IRT-based difficulty). The agreement is weighted using a linear scheme, that is, for instance, rating a task as $0$ and $1$ will be considered as a better agreement than $0$ and $2$. This results in a $\kappa = 0.21$ for \textit{HumanEval+} and a $\kappa = 0.17$ for \textit{ClassEval}, resulting in low agreements, respectively a Slight and Fair agreement \cite{Munoz97}. As such, this points toward human annotators not being a reliable proxy for the difficulty as estimated by CodeLLMs, as there exists a non-trivial number of tasks on which both sides disagree. While Alhamed et al. \cite{Alhamed21} showed that annotators tend to underestimate the real difficulty of the tasks, we found that this does not affect our result. Indeed, even if we were to increase the difficulty reported by human annotators by one category (i.e., accounting for the human annotators underestimating the real difficulty), the $\kappa$ coefficient of agreement between the human annotators and the CodeLLMs' assessment would not change significantly. We found similar results if we consider that human annotators overestimate the difficulty.

\begin{table}[]
\caption{Examples of difficulty ratings between humans and LLMs for some tasks of \textit{HumanEval+} (\textbf{Left}) and \textit{ClassEval} (\textbf{Right}). Numbers represent the reported difficulty on the scale used by our human assessors. We discretise the difficulty evaluated for our LLMs on the same scale for comparison.}
\label{tab:tasks_comp_human_llm}
\resizebox{0.95\textwidth}{!}{
\begin{tabular}
{c|c|c||c|c|c}
\multirow{2}{*}{Task Description (\textit{HumanEval+})} & \multicolumn{2}{c||}{Judgment} & \multirow{2}{*}{Task Description (\textit{ClassEval})} & \multicolumn{2}{c}{Judgment}\\
& LLMs & Humans & & LLMs & Humans\\
\toprule
Convert String to MD5 &      0     &         2     & Convert Snake Game component &     0     &          3      \\
Encoding by shifting characters       &    5      &        2     & Hexadecimal conversion &      0     &          2    \\
Prime factors of an integer       &   0      &         3     & XML files write &     1     &           3\\
Triplet under conditions       &   2     &         4     &          Parse and split url &    4     &   2 \\
List with sentences as elements       &  1
&    2      &   Convert Unicodes in HTML &     5     &     3 \\
\bottomrule
\end{tabular}
}
\end{table}

Thus, this shows that the estimation by human annotators can, at least in several cases, drastically differ from CodeLLMs. For instance, in \textit{HumanEval+}, several of the easiest tasks for CodeLLMs that were assessed as harder by participants are mathematical/logical in nature, for example, sorting tasks based on a condition (like binary representation of the numbers and prime numbers). In this case, while the tasks themselves are not inherently difficult, they become more complex due to additional concepts that human annotators perceive as increasing the difficulty. This might, of course, differ depending on the annotators, yet the ratings we aim to capture are a general assessment. In contrast, tasks with convoluted prompts or multiple conditions, which were challenging for CodeLLMs, tended to be rated simpler by humans. We give examples of such tasks in Table \ref{tab:tasks_comp_human_llm}.

In that case, CodeLLMs might struggle to keep track of all conditions. We give an example in Listing \ref{list:example-4}. In that example, CodeLLMs' difficulty is high (0.88) while human annotators' median announced time is 5 min (low difficulty). Human annotators note that the task is easy, as it is a basic string manipulation with a simple edge case. Observing CodeLLMs' codes, it is clear that the CodeLLMs struggle with handling the two conditions, especially the last one. Indeed, the CodeLLMs' code generally messes the two conditions up by merging them (e.g., \enquote{Example 1} should be \enquote{Example-1}, but the generated code generally returns \enquote{Example-\_\_1}).

\begin{lstlisting}[language=python, caption={Example of a task, represented here by a single prompt, difficult for CodeLLMs but judged simple by human-annotators.}, label={list:example-4}]
def fix_spaces(s: str) -> str:
  """Write a function named 'fix_spaces' that modifies a given string by replacing all spaces with underscores. If a string contains more than two consecutive spaces, these spaces should be replaced with a single dash."""
  
\end{lstlisting}

\begin{tcolorbox}[colback=blue!5,colframe=blue!40!black]
\textbf{Findings 4:} Human annotators' assessments differ from CodeLLMs' difficulty assessment, emphasising that using human judgement to assess CodeLLMs' task difficulty might not be a valid proxy. This motivates the usage of approaches like ours that rely directly on CodeLLMs' assessments to establish characteristics of the tasks.
\end{tcolorbox}

\section{Discussion}\label{sec:discussion}

With TaskEval, we showed how to leverage IRT’s outcome to analyse benchmarks to provide a more fine-grained analysis. While the aim of our study was not to primarily focus on addressing practical aspects such as computational cost or scalability, we highlighted several key practical points and considerations that could benefit practitioners.

\textbf{Multi-prompt evaluation:} Echoing previous studies, we re-emphasise that multiple prompts are a necessity to assess and evaluate CodeLLMs \cite{anwar2024foundationalchallengesassuringalignment} properly. Indeed, variation in the prompts can drastically alter the output as we saw, for instance, in the motivating examples in Section \ref{sec:motiv} with the Pass@1 drastically changing depending on the rephrasing and/or level of information in the prompt. In our case, this was even more important as we aimed to evaluate concepts at the task-level, i.e., as independent of the formulation of the prompt as possible. Beyond evaluation, from a user perspective, extra care needs to be taken when prompting a CodeLLM, especially if the output of a single generation will be used. Using prompt tuning methods can help eliminate some biases \cite{fernando2023promptbreederselfreferentialselfimprovementprompt}, but it may not completely address them. As such, a trade-off should be reached between using the output of a single prompt and using too many to be leveraged effectively by a user. In that regard, the use of differential testing or partial oracles \cite{wang2024validatingllmgeneratedprogramsmetamorphic} could be a promising venue of improvement.

\textbf{Human assessment of difficulty:} We showed that difficulty as assessed by humans might not be a valid proxy for CodeLLMs, and so a misalignment between human and CodeLLMs task difficulty might exist. Thus, using difficulty as reported in code-games benchmarks, such as CodeContest \cite{doi:10.1126/science.abq1158}, might not apply when evaluating CodeLLMs. Leveraging instead the difficulty as computed by an IRT model from the CodeLLMs’ outputs themselves might be more truthful to the real difficulty of the task for CodeLLMs. This observation also sheds light on a broader cautionary point: since benchmarks are often used to assess CodeLLM capabilities, potentially in relation to developers’ skills, if they mischaracterise what is simple or difficult for a CodeLLM, claims about CodeLLMs might be overstated.

\textbf{Identifying CodeLLMs shortcomings}: In our study, analysing tasks through the prism of IRT allows for some insights into the benchmark under study. Identifying tasks with varying difficulty helps reveal potential shortcomings of CodeLLMs, especially when this concerns a whole topic, as we saw in RQ2. From there, further analysis can be conducted to understand why a difficulty is present. This could potentially be automated to provide a streamlined process of identifying shortcomings or hard topics. Then, the shortcomings should be addressed, potentially by fine-tuning the CodeLLMs on similar tasks/certain prompt templates or exploiting in-context learning (such as few-shot learning). To further train CodeLLMs, one can generate new tasks using the extracted information with the help of an approach such as Self-Instruct \cite{SelfInstruct}, which allows generating new tasks based on a subset of sampled tasks through a generator LLM. This subset of tasks can be, for instance, the identified topics, as the prompt similarity between tasks can ease the generation. However, only giving a list of tasks marked as easy/hard is likely not to allow us to generate new hard/easy tasks accurately. Indeed, the limited number of tasks as well as the lack of explanation on why a given task is hard/easy would mean LLM tasked with generating new tasks will tend to struggle to identify what makes a task hard/easy just from the prompt. Instead, using a combinatory approach where one asks the generator LLM to augment a hard task by adding a constraint, or by adding a constraint from a hard task to an easy one (i.e., some form of fuzzing), could provide better results. This would ultimately reduce the diversity of generated tasks, but increase the likelihood that the generated tasks are hard.

\textbf{Evaluating benchmark discriminability:} Discriminant can be used to further select tasks to distinguish between CodeLLMs. Indeed, if the goal is, for a given benchmark, to differentiate between CodeLLMs, only the most discriminant tasks are required, as these tasks are the ones for which there are the most differences between CodeLLMs. On the contrary, focusing on low discriminant tasks can also highlight shortcomings, but of a different nature. In that case, different CodeLLMs will perform similarly to each other. If all CodeLLMs also struggle on these tasks, it can emphasise particular tasks for which CodeLLMs under study are not adequately aligned. For instance, in RQ2, from Figure \ref{fig:topics}, we saw that the tasks in the \textit{Condition checking} or \textit{Data structure calculation} topics of ClassEval had an average difficulty of 0.4 and 0.64 and an average discriminant of 0.65 and 1. Since all models can somehow struggle with tasks on these topics, regardless of their capability, this underscores that such tasks can still pose significant challenges for CodeLLMs.

\textbf{Cost and Scalability:} In terms of cost, the most expensive part in our experiments was prompt generation and verification. The overall cost is proportional to the number of benchmarks, tasks, and transformations applied. For the prompt generation, using GPT-4 turbo, this amounted to ~50\$ for all tasks and all levels ($\sim$6,500 prompts), accounting for rerun when correcting them, resulting in 0.007\$/prompt. Note that GPT-4 turbo is an older model and the API cost has since decreased drastically on newer versions, which could easily divide the cost of the experiments we did. For the prompt verification, this had to be done manually, which translated to several dozen hours of work. This was done to ensure as few errors as possible in the generated prompts, and is a limiting factor for the scalability that could be a future work opportunity. Note this, however, needs only to be done once for the benchmarks under study. For the generation, the cost is proportional to the number of models (8 in our case) and the number of inferences ($\sim$32,00 in our case). In our case, the open-source models used were run on a single 32GB VRAM GPU with less than an hour of runtime each. For the closed-source model (GPT3.5 and DeepSeek-V3), the cost was ~5\$ each, for all tasks and all prompts, resulting in 0.0002\$/inference. IRT model training runtime/cost is negligible. Overall, we believe the approach is relatively inexpensive in terms of cost, with only the manual verification being a limiting factor, and so could be easily scalable, either through application to new benchmarks or to more models.

\section{Related Works}\label{sec:rel}

\textit{\textbf{Evaluating CodeLLMs:}} Evaluation of CodeLLMs in the most simple form involves using a benchmark, and having a CodeLLM under test generates code for each prompt in the benchmark. Then, an aggregated metric can be calculated to evaluate the performance of the CodeLLMs on the benchmark. Several benchmarks for code have been proposed over the years, each with some particularity: besides \textit{HumanEval+} and \textit{ClassEval} used in this work, \textit{APPS} \cite{APPS} is a benchmark based on coding competition code fragments, \textit{CodeXGlue} \cite{CodeXGlue} contains additional tasks beyond code generation, such as code completion or code translation or \textit{BigCodeBench} \cite{BigCodeBench}, which contains more complex instructions. The difference between those benchmarks mainly stems from the different levels of dependency needed to solve the task (e.g., a simple function for \textit{HumanEval+} or class level for \textit{ClassEval}), yet the evaluation mainly relies on aggregation over the whole benchmark using accuracy or pass@k. More recently, multiple benchmarks are being proposed to measure a particular property of the CodeLLMs by collecting/crafting in a particular way: Coignion et al. \cite{LeetCodeSpeed} evaluated the speed of generated code based on LeetCode, Liu et al. \cite{HalluCode} created \textit{HalluCode} to evaluate CodeLLMs' attitude towards different types of hallucination, or Jain et al. \cite{Livecodebench} introduced LiveCode bench to measure contamination in CodeLLMs. However, the properties evaluated are generally at the benchmark level, and the possible impact of the rephrasing is not evaluated. Moreover, when it comes to difficulty, most benchmarks make use of human judgment, for instance, in the case of APPS, by using the subjective score given by code game contestants. We showed in RQ4 that this assessment might not translate to the actual difficulty of realising the task for CodeLLMs. This result echoes similar observations in other studies: Ouyang et al. \cite{GPTComp} showed that the test pass rate does not necessarily correlate with difficulty, as aggregated by human participants in a coding contest, and Kou et al. \cite{HumanAttention} highlighted major differences when comparing how a human developer and CodeLLMs process code instructions.

\textit{\textbf{Item Response Theory to evaluate NLP models:}} IRT \cite{baker2001basics} has been used extensively to evaluate NLP models. Sedoc et al. \cite{Sedoc20} used IRT to compare chatbots' proficiency by leveraging human annotators' judgment on the chatbots' output as observations. They notably show that the IRT model can be used to further streamline evaluations by selecting the most discriminative examples. Rodriguez et al. \cite{Rodriguez21} propose to leverage IRT to redefine leaderboards in NLP model evaluation by defining the Difficulty and Ability Discriminating (DAD) approach. They show that IRT models offer greater reliability in assessing the ranking stability of questions both within and across benchmarks and can provide valuable qualitative insights into the evaluated leaderboards. Similarly, Vania et al. \cite{Vania21} use a unique IRT model to analyse 29 NLP datasets and 18 Transformers, producing insights into which dataset contains more difficult examples or which is more effective at distinguishing strong models. Zhuang et al. \cite{Zhuang24} explain how IRT can be used for adaptive testing of CodeLLMs, highlighting the limitations of traditional metrics such as Accuracy. First, compared to those studies, we focus on coding tasks and, more particularly, code generation. Secondly, these studies consider an analysis of the benchmark prompts. On the contrary, we aim at measuring the difficulty of the \textit{task} itself, as previous studies showed that prompt formulation has a drastic impact on the models' output \cite{MultiPrompts, InstructionDiff, ReCode}. Moreover, our approach differs from prompt engineering techniques \cite{sasaki2024systematic} as they aim to tune the model/prompt to achieve the best result on a particular task. In contrast, we generate several prompts to represent the diversity of possible CodeLLM answers for a particular task and observe the result.

\section{Threats to Validity}\label{sec:threat}

\textbf{Construct validity:} To derive the characteristics of the tasks, we rely on IRT, which has been extensively used in education measurement \cite{zanon2016application}, as well as in assessing LLM in classical NLP settings \cite{rodriguez2021evaluation, lalor2016building}. IRT assumes local independence, which might be jeopardised in our case because we use multiple prompts for the same task. To avoid a potential local dependence effect, we aggregate results of prompts for the same task, effectively building polytomous superitems or \enquote{testlets} \cite{Wainer1987item, Wainer_Bradlow_Wang_2007}. This was shown to alleviate the local dependence effect. As such, the backbone of our approach is grounded in theory. We used a multi-prompted approach as previous studies highlighted the importance of multiple phrasings. We rely on three levels of prompts to model the effect of information on how the CodeLLMs handle the task. The choice of three levels was made based on our observations. So, while we limited our study to three levels of prompts, we do not expect that providing more information in prompts will drastically affect our results. Another threat might come from the generation of the prompts. The transformations used are motivated based on existing works. Moreover, while we have made use of state-of-the-art GPT-4 to generate them, we manually checked the prompts to make sure that the generative process did not alter semantics. Choosing GPT4 for prompt generation might unintentionally bias the evaluation by benefiting a model trained similarly (GPT3.5). However, observing our results, especially the reported abilities by the trained IRT models, seems to point towards this not influencing the results. Indeed, the abilities obtained reflect the \enquote{perceived} capacity of the model. Finally, we did not consider examples in the prompts. This, as we mentioned, can decrease the performance of the models and so add artificially hard tasks. This was nonetheless necessary to make the comparison fair across tasks, as not all examples contain the same information, as we gave some examples in Section \ref{sec:motiv}. However, beyond modifying some tasks' characteristics, it does not affect the applicability of \textit{TaskEval}. 

\textbf{Internal validity}: The CodeLLMs and benchmarks selected could be a threat to validity. We made sure to select different architectures through $8$ different CodeLLMs to encompass different models. We used the implementation and models' weights as present on HuggingFace \cite{huggingface} by the original authors of each model or the web API in the case of GPT3.5 and DeepSeek-V3. For the benchmarks, we selected two different datasets, both acting at different levels (functional for \textit{HumanEval+} and class for \textit{ClassEval}). Regarding our experiments, the modelled topics as well as the human-annotation data could be a limitation. For the topics, we used BERTopics to automatically label the tasks, with noisy tasks being discarded. This could reduce our potential topics list, but this allowed for an automatic process. We still obtained over 15 topics with at least $3$ tasks each. We manually reviewed the obtained topics. Regarding human annotations, the data primarily reflect how difficult annotators perceive each task to be. As detailed in RQ4, we did not ask crowd-workers to code the tasks, as this could introduce biases due to the uncontrolled use of CodeLLMs. Instead, we followed an existing approach \cite{Alhamed21} based on effort estimation, which mitigates this issue. Although the difficulty is then assessed based on developers' subjective judgment, the used approach allows for obtaining an estimation of a consensus, which limits potential outliers due to different experiences, relative estimation, etc. Moreover, even under variations due to human judgment, participants' experience, etc., we still show there is a delta between humans' and CodeLLM's perception of the difficulty, which further motivates proposing an approach that is LLM-centric.

\textbf{External validity}: Our selection of tasks from the chosen benchmarks may pose a threat to validity, as the collection might not fully represent real-world programming tasks for assessing LLMs' capabilities. To mitigate this, we used two benchmarks, HumanEval+ and ClassEval, which have been used to evaluate CodeLLM performance \cite{EvalPlusLeaderboard, ClassEvalLeaderboard} in previous studies. Both datasets cover different functionality levels (function for HumanEval+ and class for ClassEval). We chose their Python version as it is the most widespread programming language when it comes to LLMs \cite{jiang2024surveylargelanguagemodels}. Finally, this study aimed to establish a foundation for assessing task characteristics, as well as showing that interesting insights can be gathered from the analysis. The approach itself should not be affected by more practical benchmarks. Nonetheless, this limits the generalisation of our study in terms of practical insights. We had to limit ourselves to these two datasets to reduce the complexity of both the analysis and the cost of the prompt generation/human judgement experiments. Studying more advanced benchmarks (e.g. SWE-bench \cite{jimenez2024swebench}), benchmarks using other programming languages (e.g. Multi-Eval \cite{mbxp_athiwaratkun2022}) or industry-oriented benchmarks would add valuable insights to the capacity of LLMs on a more diverse panel of tasks, and so is an important future venue of research.

\textbf{Reliability validity}: We provide a detailed description of our methodology and make our code and data publicly available to facilitate reproducibility and further research \cite{rep-package}.
 

\section{Conclusion}\label{sec:conclusion}
In this paper, we present a method for analysing the characteristics of code generation tasks for CodeLLMs. Specifically, we introduce \textit{TaskEval}, a novel framework designed to assess and explore task characteristics in depth. We evaluated \textit{TaskEval} on two well-known benchmarks for CodeLLM-based code generation: \textit{HumanEval+} and \textit{ClassEval}. A set of diverse prompts is crafted for each task, including different levels of contextual information about the task and various rephrasings, accounting for the impact of prompt variations on the generation. Then, eight CodeLLMs were used to solve coding tasks using the set of prompts. \textit{TaskEval} is based on Item Response Theory and allows for computing difficulty (i.e., how hard is a given task for CodeLLMs) and discriminant (i.e., how do CodeLLMs compare to each other on a given task) of each task within a benchmark. Using our method, we conducted multiple analyses on the tasks, namely, topic-wise, program construct-wise, as well as a comparison with human annotators. Based on thematic similarities, we clustered tasks from both benchmarks into 17 and 21 topics to inspect their characteristics, revealing topics with high difficulty and high discriminant for CodeLLMs (e.g., \textit{Sequences}). Similarly, we assess program construct (e.g., conditions, variable assignment, etc.) given the tasks' difficulty/discriminant and highlighted discrepancies/similarities between benchmarks. Finally, we compared the difficulty assessment between CodeLLMs and human annotators, revealing stark differences in the assessment. In future work, we plan to extend the evaluation of \textit{TaskEval} to additional domains, such as code completion and summarisation. Furthermore, we aim to build on our analysis to enhance task selection (e.g., based on discriminative features) and facilitate task generation (e.g., using topic modelling).

\section*{Acknowledgments}

This work was partly supported by: Fonds de Recherche du Québec (FRQ), the Canadian Institute for Advanced Research (CIFAR) as well as the DEEL project CRDPJ 537462-18 funded by the 
Natural Sciences and Engineering Research Council of Canada (NSERC) and the Consortium for Research and Innovation in Aerospace in Québec (CRIAQ), together with its industrial partners Thales Canada inc, Bell Textron Canada Limited, CAE inc and Bombardier inc.

\bibliographystyle{ACM-Reference-Format}
\bibliography{sample-base}

\end{document}